\documentclass[aps,prd,twocolumn,tightenlines,preprintnumbers,showpacs,superscriptaddress,notitlepage,nofootinbib,eqsecnum,floatfix,
  bibnotes,
  10pt]{revtex4-1}

\usepackage{graphicx}  
\usepackage{dcolumn}   
\usepackage{bm}        
\usepackage{amssymb}   
\usepackage{standalone}
\usepackage{enumitem}
\usepackage[pdftex]{color}
\usepackage[utf8]{inputenc}
\usepackage{xcolor}
\usepackage{slashed}
\usepackage{booktabs}
\usepackage{multirow}
\usepackage{amsmath}
\usepackage{bbm}
\usepackage{bbold}
\usepackage{stackrel}
\usepackage{rotating}
\usepackage{CJKutf8}
\usepackage{pifont}
\usepackage{mathtools}
\usepackage[normalem]{ulem}

\usepackage{hyperref}
\hypersetup{
    colorlinks=true,       
    linkcolor=blue,          
    citecolor=blue,        
    filecolor=blue,      
    urlcolor=blue           
}
\usepackage{simplewick}

\usepackage{float} 
\usepackage{tikz} 
\usepackage{xspace}

\newcommand{\mc}[1]{\ensuremath{\mathcal{ #1 }}}
\newcommand{\matrixel}[3]{\left< #1 \vphantom{#2#3} \right| #2 \left| #3 \vphantom{#1#2} \right>}
\newcommand{\braket}[2]{\left< #1 \vphantom{#2} \right| \left. #2 \vphantom{#1} \right>}
\newcommand{\avg}[1]{\left< #1 \right>}
\newcommand{\abs}[1]{\left| #1 \right|}
\newcommand{\cpthreemath}{
    CP$^3$-Origins, Dept. of Mathematics and Computer Science,
    University of Southern Denmark, 
    5230 Odense M, Denmark
}
\newcommand{\cpthreephys}{
    CP$^3$-Origins, Dept. of Physics, Chemistry and Pharmacy,
    University of Southern Denmark, 
    5230 Odense M, Denmark
}

\newcommand{\dias}{
    Danish Institute for Advanced Study, 
    University of Southern Denmark,  5230 Odense M, Denmark
}

\newcommand{\napoli}{
    Dipartimento di Fisica “E. Pancini", 
    Università di Napoli Federico II - INFN sezione di Napoli, 
    Complesso Universitario di Monte S. Angelo, 80126 Napoli, Italy
}

\newcommand{\uoe}{
    School of Physics and Astronomy,
        The University of Edinburgh,
    EH9 3FD Edinburgh, United Kingdom
    }
\newcommand\cern{CERN, Theoretical Physics Department, 1211 Geneva 23, Switzerland}

\begin{document}

\title{The spectrum of QCD with one flavour: A Window for Supersymmetric Dynamics}
\author{Michele~Della~Morte}
\affiliation{\cpthreemath}
\author{Benjamin~J{\"a}ger}
\affiliation{\cpthreemath}\affiliation{\dias}
\author{Francesco Sannino}
\affiliation{\dias}\affiliation{\cpthreephys}\affiliation{\napoli}
\author{J.~Tobias~Tsang}\thanks{Corresponding author}\email{j.t.tsang@cern.ch}
\affiliation{\cern}\affiliation{\cpthreemath}
\author{Felix~P.~G.~Ziegler}
\affiliation{\uoe}
\begin{abstract}
We compute the spectrum of the low-lying mesonic states with vector, scalar and
pseudoscalar quantum numbers in QCD with one flavour. With three colours the
fundamental and the two-index anti-symmetric representations of the gauge group
coincide. The latter is an orientifold theory that maps into the bosonic sector
of $\mc{N}=1$ super Yang-Mills theory in the large number of colours limit.

We employ Wilson fermions along with tree-level improvement in the gluonic and
fermionic parts of the action. In this setup the Dirac operator can develop real
negative eigenvalues. We therefore perform a detailed study in order to identify
configurations where the fermion determinant is negative and eventually reweight
them.  We finally compare results with effective field theory predictions valid
in the large $N_C$ limit and find reasonably consistent values despite $N_C$
being only three. Additionally, the spin-one sector provides a novel window for
supersymmetric dynamics.
\end{abstract}

\maketitle
\preprint{CERN-TH-2023-028}

\tableofcontents

\section{Introduction \label{sec:intro}}
Understanding the dynamics of strongly coupled gauge theories, such as QCD, has
motivated the construction of several expansions complementary to the standard,
perturbative, weak coupling expansion.  One of the most prominent examples is
the large $N_C$ limit (where $N_C$ is the number of colours), introduced by
't~Hooft in Ref.~\cite{tHooft:1973alw}.  In this case one keeps quarks in the
fundamental representation of the gauge group $SU(N_C)$ and organises an
expansion in $1/N_C$ using a diagrammatic approach. Several properties of QCD
can then be understood in a simple way, suggesting that $N_C=3$ is ``large''.
However, since quark loops are suppressed in this expansion, the properties of
the $\eta'$-meson are not well reproduced in the 't~Hooft large $N_C$
limit. Baryons also become increasingly heavy as $N_C$ grows.

Partly motivated by that, Corrigan and Ramond (CR) introduced a different large
$N_C$ expansion in Ref.~\cite{Corrigan:1979xf}, in which quarks transform
according to the two-index antisymmetric representation of the gauge
group. While 't~Hooft and CR expansions coincide for $N_C=3$, they are very
different in the large $N_C$ limit. Notably, in the CR expansion, quark loops
are not suppressed as $N_C \to \infty$.  A simple scaling of the dimensionality
of the representations of the quark fields suggests that the CR large $N_C$
limit may share non-trivial dynamical properties with supersymmetric theories.
This relation has been made precise by Armoni, Shifman and Veneziano in
Refs.~\cite{Armoni:2003gp,Armoni:2003fb}, where a connection between the mesonic
sectors of the two-index (anti-)symmetric theories and of $\mathcal{N}=1$ super
Yang-Mills theory (sYM) is established. The subtle issues of the confinement
properties and (in)equivalences at large $N_C$ were investigated in
Ref.~\cite{Sannino:2005sk}.  Further developing the correspondence, in
Ref.~\cite{Sannino:2003xe} supersymmetry inspired effective Lagrangians have
been constructed for gauge theories featuring one Dirac fermion transforming
either in the symmetric or in the anti-symmetric two-index representation of the
gauge group $SU(N_C)$ (orientifold theories).  At leading order in the $1/N_C$
expansion such effective theories coincide with that of supersymmetric
{gluodynamics} restricted to its mesonic sector.  These correspondences imply
that non-perturbative quantities computed in orientifold theories can be
related, up to $1/N_C$ effects, to the analogous ones in sYM.  By considering
$1/N_C$ supersymmetry breaking effects, including the explicit ones due to a
finite quark mass, a number of predictions are made in
Ref.~\cite{Sannino:2003xe} concerning the spectrum of the low-lying mesonic
states.\footnote{A string theory dual of orientifold theories has
also been used in Ref.~\cite{Armoni:2005qr} to make predictions in the
massless limit.}  In this work we confront such predictions with non-perturbative results
produced by means of lattice simulations. For simplicity, in this first study we
only consider $N_C=3$, which corresponds to one-flavour QCD. This has the
advantage that available simulation packages for lattice QCD can be used without
having to develop new code for handling representations of the fermionic fields
different from the fundamental one.  Future studies will be devoted to the
extension to $N_C>3$. Intriguingly, by flipping the point of view
(cf. Refs.~\cite{Sannino:2005sk,Feo:2004mr}), we can use QCD results to learn
about the spectrum and dynamics of supersymmetric theories, in particular
$\mathcal{N} = 1$ sYM. Analytic and numerical studies can now be
employed to investigate several dynamical properties, including the
theta-angle~\cite{Sannino:2003xe}.

One-flavour QCD has been the object of several previous lattice studies. The
qualitative behaviour of the theory has been discussed in
Ref.~\cite{Creutz:2006ts}. In Ref.~\cite{DeGrand:2006uy} the quark condensate
has been computed by comparing the density of low-lying eigenvalues of the
overlap Dirac operator to predictions from Random Matrix
Theory~\cite{Leutwyler:1992yt,Shuryak:1992pi}. The result is consistent with the
prediction for the gluino condensate in sYM obtained in
Ref.~\cite{Armoni:2003yv}.  Using Wilson fermions, Ref.~\cite{Farchioni:2007dw}
presents a computation of the low-lying hadronic spectrum of one-flavour QCD. We
improve here on that computation by considering a finer lattice spacing, larger
volumes and a tree-level improved fermionic action. In
Ref.~\cite{Francis:2018xjd} the one-flavour $SU(2)$ vector gauge theory with the
fermion in the fundamental representation is studied as a possible composite
model for Dark Matter (DM). The Dirac operator is discretised using Wilson's
regularisation. The fundamental representation of $SU(2)$ is pseudo-real making
the global symmetries and dynamics different from three colours QCD.  In
particular, the dark-matter model of Ref.~\cite{Francis:2018xjd} features a
mass-gap with vector mesons being the lightest triplet of the enhanced $SU(2)$
global symmetry. A similar DM model based on $SU(2)$ gauge theory with scalar
quarks was proposed in Ref.~\cite{Hambye:2009fg}.

Finally, in Ref.~\cite{Athenodorou:2021wom} the single flavour $SU(2)$ theory is
considered with the fermion in the adjoint representation.  The goal in this
case is to gain insights on the emergence of the conformal window. Again the
Wilson Dirac operator is used in the numerical simulations.  As is highlighted
by this brief review, one-flavour QCD is implemented on the lattice by adopting
either overlap (or more generally Ginsparg-Wilson) or Wilson fermions.  That is
because in those cases the single-flavour lattice Dirac operator can be
rigorously defined.  Wilson fermions are computationally cheaper but in such
regularisation the spectrum of the Dirac operator may contain real negative
eigenvalues for positive (but small) quark masses.  That might cause a sign
problem as the fermion determinant may become negative on some configurations.
Following
Refs.~\cite{Edwards:1997sp,Edwards:1998sh,Akemann:2010em,Mohler:2020txx} we
discuss in detail how we monitor such cases.\footnote{An alternative approach
  relying on the Arnoldi algorithm to compute the eigenvalues of the
  non-Hermitian Wilson Dirac operator has been introduced in
  Ref.~\cite{Bergner:2011zp}.}

Earlier numerical investigations of orientifold
theories~\cite{Lucini:2010kj,Armoni:2008nq} used the quenched approximation
where the sign problem is absent.

Directly simulating supersymmetric gauge theories on the lattice has been an
active research field for many years. Since the literature is vast we refer the
reader to the recent review in Ref.~\cite{Schaich:2022xgy} and references within
for a discussion of the status and open problems.

A preliminary account of the results we present in this paper appeared in
Refs.~\cite{Ziegler:2021nbl,DellaMorte:2022htz}. The latter in particular
contains some algorithmic exploratory studies for $N_C=4,5$ and $6$.

The remainder of this paper is organised as follows.  In Section~\ref{sec:ens}
we describe our computational setup and provide algorithmic details.  In
Section~\ref{sec:evs} we investigate the consequences of the sign problem in our
simulations.  In Section~\ref{sec:cors} we report on the correlation function
fits required to extract the spectrum at non-zero quark masses, before
extrapolating the meson spectrum to vanishing quark masses in
Sec.~\ref{sec:results}. Finally, in Section~\ref{sec:discussion} we confront the
effective field theory predictions with our results and provide an outlook.

\section{Simulation setup \label{sec:ens}}

For the gauge part of the action, we employ the Symanzik improved gauge
action~\cite{Iwasaki:1983iya} with a fixed value for the gauge coupling of
$\beta = 4.5$. As fermion action we use one flavour of tree-level improved
Wilson fermions~\cite{Sheikholeslami:1985ij} and set the parameter of the clover
term to 1. The Wilson-Dirac operator $D$ in clover improved form is defined as
follows
\begin{align}
D(m_0) =& \frac{1}{2} \sum_{\nu = 0}^3 \left(
\gamma_{\nu} (\nabla_{\nu}^* + \nabla_{\nu})
- a \nabla_{\nu}^* \nabla_{\nu}\right)\nonumber \\[2ex]
& +  \ ac_{\mathrm{SW}} \sum_{\nu,\rho = 0}^3 \frac{i}{4} \sigma_{\nu \rho} \hat F_{\nu \rho} + m_0
\,,
\label{eq:Wilson-Dirac-operator}
\end{align}
where $a$ is the lattice spacing, $m_0$ is the bare quark mass and $\nabla_{\nu}^{(*)}$ denotes the
covariant forward (backward) derivative.  The hopping parameter $\kappa$ is
related to the bare mass $m_0$ by $1/\kappa~=~2(am_0 + 4)$.

\begin{table}
\centering
\begin{tabular}{ccccccc}
\hline
$L/a$ & $\kappa$ & ML steps & $\tau_{MD}$ & $\Delta \text{cfg}$ [MDU]& $N_\text{config}$ & Acceptance \\
\hline\hline
12 & 0.1350 & 1,1,6 & 2.0 & 64 & 877 & 0.998 \\
12 & 0.1370 & 1,1,6 & 2.0 & 64 & 778 & 0.997 \\
12 & 0.1390 & 1,1,6 & 2.0 & 64 & 731 & 0.996 \\
12 & 0.1400 & 1,1,6 & 2.0 & 64 & 674 & 0.996 \\
\hline
16 & 0.1350 & 1,1,8 & 3.0 & 120 & 1512 & 0.999 \\
16 & 0.1370 & 1,1,8 & 3.0 & 120 & 539 & 0.998 \\
16 & 0.1390 & 1,1,8 & 3.0 & 120 & 1189 & 0.997 \\
16 & 0.1400 & 1,1,8 & 3.0 & 120 & 959 & 0.994 \\
16 & 0.1405 & 1,1,8 & 3.0 & 120 & 686 & 0.991 \\
16 & 0.1410 & 1,1,10 & 3.0 & 120 & 989 & 0.957 \\
\hline
20 & 0.1350 & 1,1,6 & 2.0 & 64 & 503 & 0.996 \\
20 & 0.1370 & 1,1,6 & 2.0 & 64 & 180 & 0.993 \\
20 & 0.1390 & 1,1,8 & 3.0 & 120 & 346 & 0.993 \\
\hline
24 & 0.1350 & 1,1,10 & 2.0 & 64 & 360 & 0.999 \\
24 & 0.1390 & 1,1,6 & 2.0 & 64 & 324 & 0.986 \\
24 & 0.1405 & 1,1,6 & 2.0 & 64 & 286 & 0.966 \\
24 & 0.1410 & 1,1,9 & 2.0 & 64 & 593 & 0.841 \\
\hline 
32 & 0.1390 & 1,1,6 & 2.0 & 64 & 180 & 0.979 \\
32 & 0.1400 & 1,1,6 & 2.0 & 64 & 376 & 0.967 \\
\hline
\end{tabular}

    \caption{Overview of the lattice ensembles generated in this study. All
    configurations are at a fixed gauge coupling of $\beta=4.5$ and a fixed
    temporal extent of $T/a = 64$.  The simulation parameters were tuned to
    achieve a high acceptance with a large trajectory length $\tau_{MD}$. We
    refer the reader to the text for the definitions of the parameters.
    } \label{tab:ens}
\end{table}

In order to map out the relevant parameter space we generated 19 gauge field
ensembles covering different hopping parameters $\kappa$ between 0.1350 and
0.1410 and volumes ranging from $12^3\times 64$ to $32^3\times 64$. An overview
of the simulation parameters can be found in Table~\ref{tab:ens}.

We measure the topological charge $Q$ by integrating the Wilson
flow~\cite{Luscher:2010iy} using a third-order Runge-Kutta scheme with a
step-size of $\epsilon = 0.01$ and $1600$ integration steps. The topological
charge at the largest flow time ($t/a^2=16$) is shown for all ensembles in
Fig.~\ref{fig:topo} in Appendix~\ref{app:ensembles}. The topological charge
behaves as expected: its distribution is narrower for lighter quark masses and
broader for larger volumes~\cite{Leutwyler:1992yt}. The Wilson flow further
allows us to estimate the lattice spacing (via the reference flow scale $t_0$)
by studying the Yang-Mills gauge action density as a function of
flow-time~\cite{Luscher:2010iy}. Since our goal is to determine dimensionless
quantities, we only quote the lattice spacing in order to enable qualitative
comparison with other lattice calculations. As there is no reference scale for a
single flavour ($N_f = 1$), we use the average of $t_0$ from
$N_f=0$~\cite{Luscher:2010iy} and $N_f=2$~\cite{Bruno:2013gha} as an estimate
for the lattice spacing with $N_f=1$. In practice, we use a value of $\sqrt{8
t_0} = 0.45\,$fm. This allows us to obtain an indicative value for the lattice
spacing of $a \approx 0.06\,$fm.

All configurations are generated using the \texttt{openQCD} software
package~\cite{openQCD}. Since we only simulate a single fermion in the sea, it
is necessary to use the rational hybrid Monte Carlo (RHMC)
algorithm~\cite{Clark:2006fx}. In the rational approximation we adopt a
Zolotarev functional of degree 10. In the absence of prior knowledge about the
optimal Zolotarev approximation -- in particular for just one flavour -- we
choose a conservative range of $0.002$ and $9.0$ as a lower and upper bound for
the position of the poles. In comparison with Ref.~\cite{Mohler:2020txx} this is
a rather loose approximation, which is relevant for the tunnelling between
regions of configuration space with positive and negative determinants of the
Dirac operator. In addition, we include frequency splitting, i.e. we factorise
the Zolotarev rational into two terms, where the first factor contains the poles
$1$ to $5$ and the second term the contribution from poles $6$ to
$10$. Throughout the entire generation, we adopt three levels of integration
schemes. The outermost employs a second-order Omelyan
integrator~\cite{OMELYAN2003272} with $\lambda = 1 / 6$, which is used for the
contributions from poles $6$ to $10$. For the inner two levels we use
fourth-order Omelyan integrators, where the remaining fermion force is
calculated in the second, and the gauge forces in the innermost level.  We tune
the number of fermion integration steps (ML steps) in the different levels to
achieve a high acceptance (between $84\%$ and $99.9\%$,
c.f. Table~\ref{tab:ens}).  The pseudofermion actions and forces are obtained
using a simple multi-shift conjugate gradient solver. For ensembles with a
lighter quark mass, i.e. with larger values of $\kappa$, we take advantage of
the deflated SAP~\cite{Luscher:2007es,Luscher:2007se} preconditioned solver
given in the \texttt{openQCD} framework. The trajectory lengths of our ensembles
are typically between 2 and 3 molecular dynamic (MD) units. In our analysis, we
use every 32nd (or 40th) trajectory, which implies that configurations are at
least 64 MD units apart from each other. For each ensemble the resulting number
of configurations $N_\mathrm{config}$ on which we perform all measurements is
listed in Table~\ref{tab:ens}.  To increase the amount of statistics and to
utilise smaller computing resources more efficiently, we branch our simulation
stream into multiple replicas after thermalisation is reached.

\begin{figure}
   \includegraphics[width=\columnwidth]{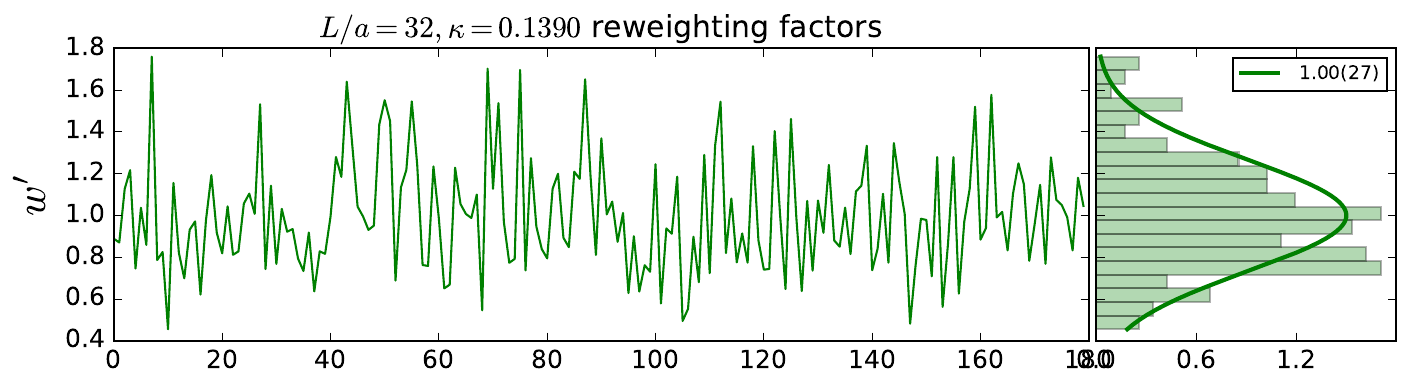}
   \includegraphics[width=\columnwidth]{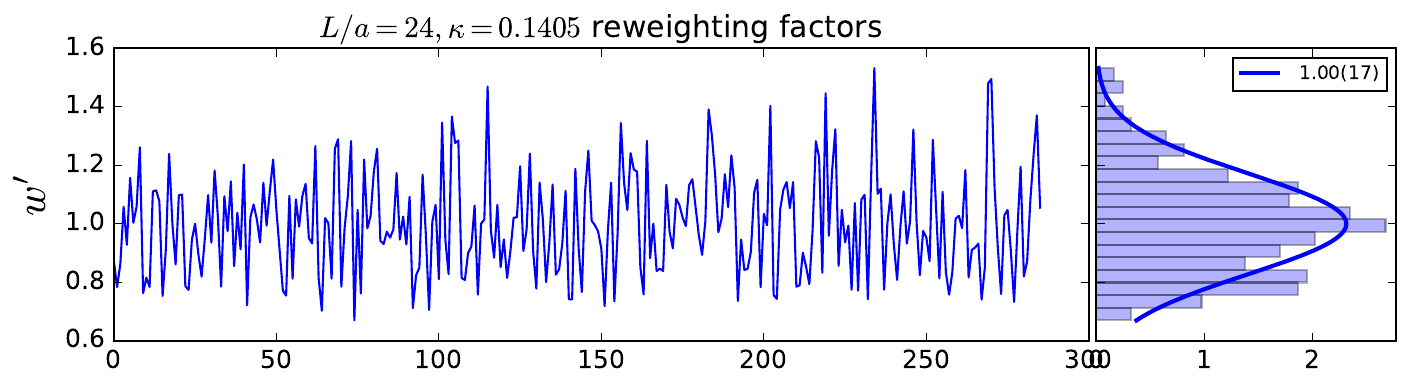}
   \caption{Normalised reweighting factors on two example ensemble.}
   \label{fig:revs_L32_k1390}
\end{figure}

\begin{figure}
   \includegraphics[width=\columnwidth]{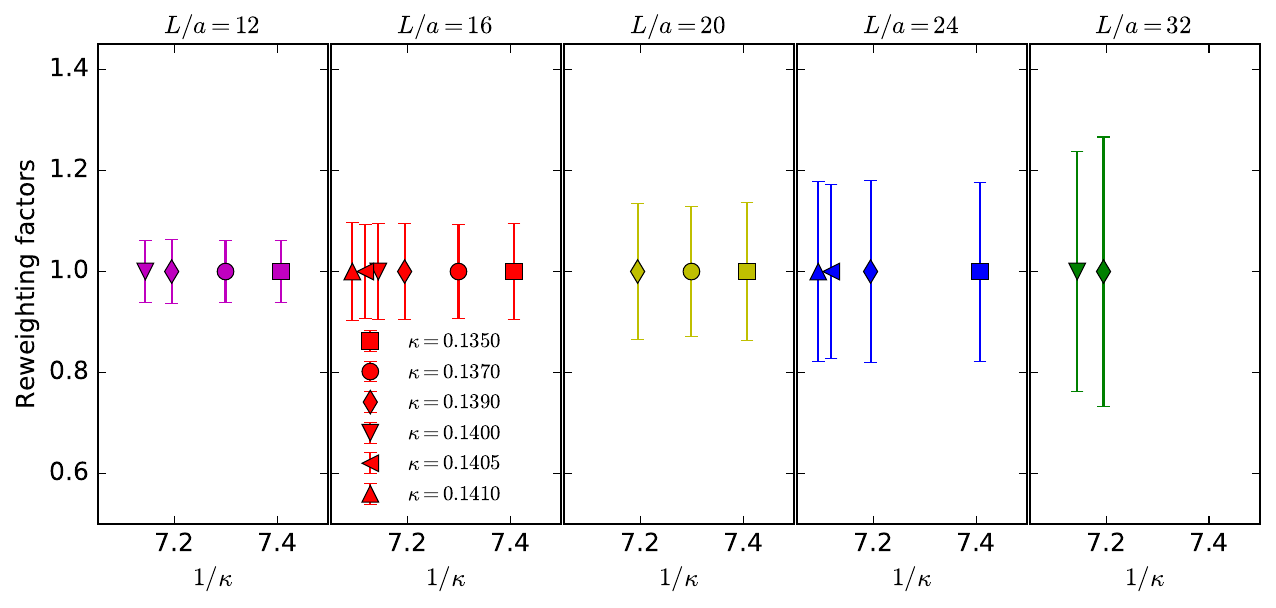} \caption{Typical
   spread of normalised reweighting factors as a function of volume and quark
   mass.}  \label{fig:revs_all}
\end{figure}

Since the Zolotarev approximation in the RHMC is not exact, we correct our
observables by using a reweighting scheme. To achieve this, on each
configuration we compute four estimators for the reweighting factors $w_i$ using
code from the \texttt{openQCD} package. The correctly reweighted gauge average
of an observable $O$ is then given by
\begin{equation}
   \avg{O}_\mathrm{rew} = \frac{\avg{w O}}{\avg{w}} = \avg{w' O}\,,
\end{equation}
where we define $w' = w/\avg{w}$. Figure~\ref{fig:revs_L32_k1390} shows these
normalised reweighting factors $w'$ as a function of the trajectory length
(excluding any thermalisation times) for two representative ensembles ($L/a=32$,
$\kappa = 0.1390$ and $L/a=24$, $\kappa = 0.1405$). In Fig.~\ref{fig:revs_all}
we show the variation of the reweighting factors for all ensembles and observe
that the fluctuations increase with volume, but are insensitive to the quark
mass.

As the phase space of this theory in the regularisation we have chosen is a
priori unknown, we computed the trace of the Polyakov loop. We find that the
Polyakov loop vanishes within errors on each ensemble, which indicates that we
are simulating in the confined phase.

\section{Eigenvalue analysis \label{sec:evs}}
\begin{figure*}
   \centering
   \includegraphics[width=.44\textwidth]{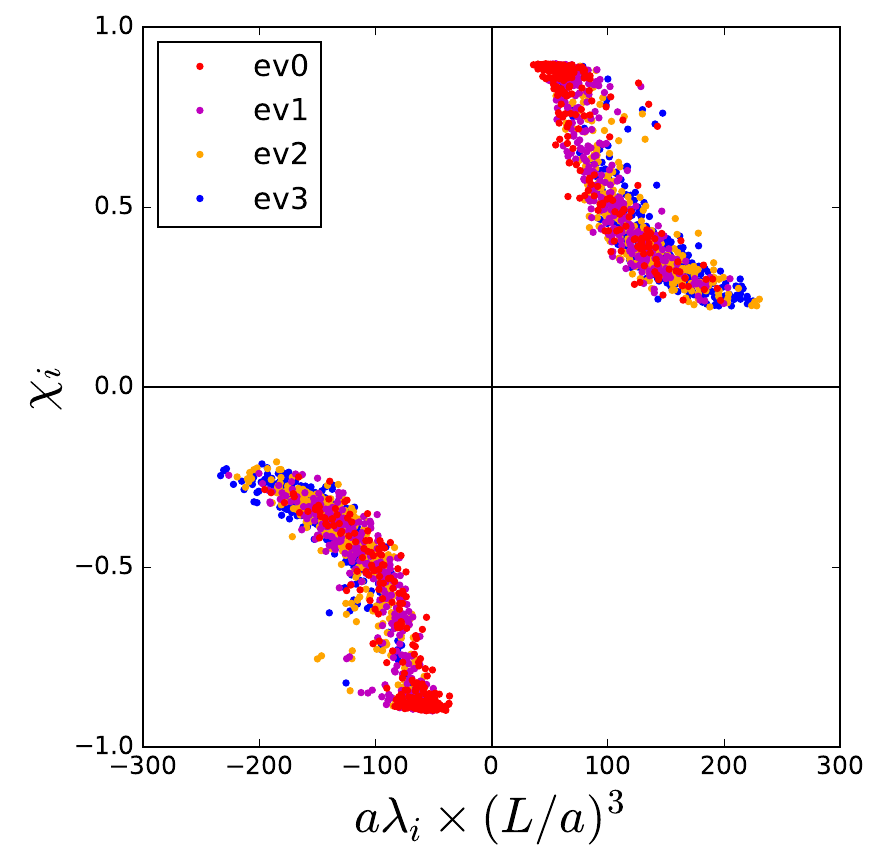}
   \includegraphics[width=.44\textwidth]{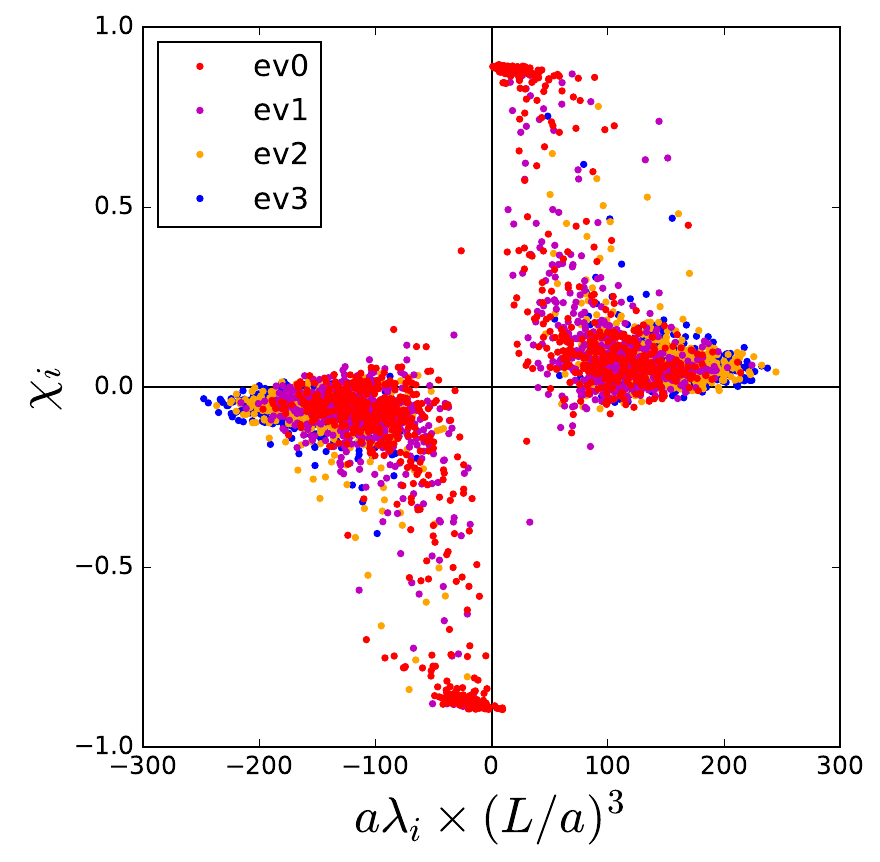}   
   \caption{Scatter plot of the lowest four eigenvalues and chiralities for
     $L/a=16$ and $\kappa=0.1405$ (left) and $\kappa=0.1410$ (right).}
\label{fig:ev-butterfly}
\end{figure*}

\begin{figure}
\includegraphics[width=\columnwidth]{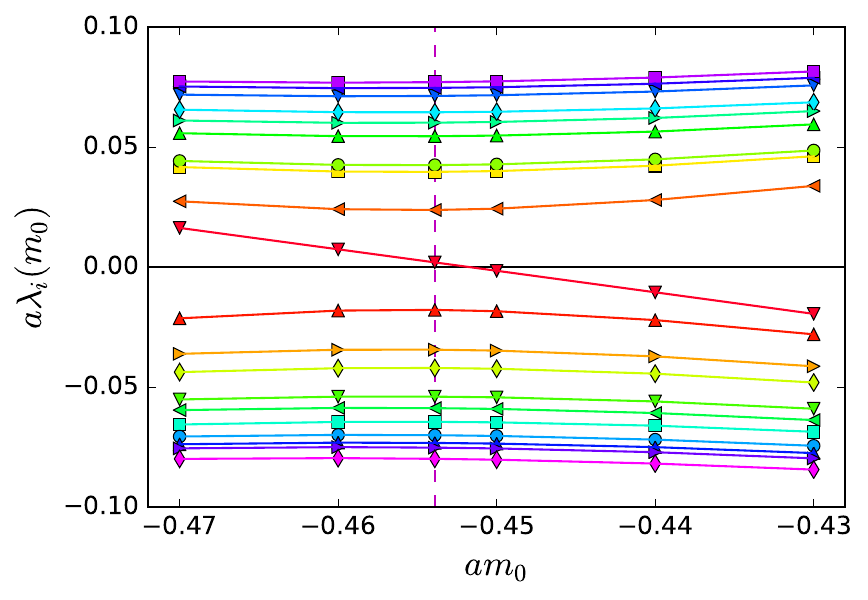}
\caption{Tracking analysis of the lowest 20 eigenvalues on a $L/a=16$, $\kappa =
  0.1410$ configuration with a negative fermion determinant.}
\label{fig:ev-tracking}
\end{figure}

The use of Wilson fermions for lattice QCD with an odd number of quark flavours
or with non-mass-degenerate (light) quarks can introduce a sign problem. This
occurs because the configuration space is divided into two sectors, one
associated to a positive sign of the fermion determinant and one to a negative
sign. These sectors are separated by a zero of the fermionic measure.  Note that
the latter translates into a pole of the fermionic force in the molecular
dynamics algorithm. With exact integration and an exact expression for the
square root function, the negative sector cannot be reached from the positive
one. In practice the algorithmic choices for the rational approximation yield a
finite (rather than infinite) barrier between the two sectors.

In the thermodynamic and continuum limit the trajectory is expected to be
constrained to the positive sector. However, at finite volume, the presence of
the negative sector has to be accounted for by sign reweighting which requires
knowledge of the sign of the fermion determinant $\det(D)$. A direct computation
is numerically (prohibitively) expensive.  Instead we follow a strategy in which
the sign of $\det(D)$ is inferred from computing a few of the lowest eigenvalues
of the Dirac operator. This can be achieved at a cost linear in the lattice
volume and using the approach we will now sketch:

Due to $\gamma_5$-Hermiticity of the Wilson-Dirac operator, i.e. 
\begin{equation}
 D^{\dagger} = \gamma_5 D \gamma_5\,,
\label{eq:g5-herm}
\end{equation}
the matrix $Q= \gamma_5 D$ is Hermitian and its spectrum is real. Furthermore,
it holds that $\det(D) = \det(\gamma_5) \det(D) = \det(Q)$ and that a zero
eigenvalue of $D$ is also a zero eigenvalue of $Q$. Recalling that the
eigenvalues of $D$ come in complex conjugate pairs, for $\det(D)$ to be negative
there must be an odd number of negative real eigenvalues of $D$.

Since the fermion determinant $\det(D)$ is assumed to be positive for large
quark masses, we can infer that the determinant at the unitary mass $m_0^*$,
used in the actual simulation, is negative if and only if there is an odd number
of eigenvalues that cross zero as the mass is decreased from large quark masses
to $m_0^*$. The idea is to locate (on each gauge configuration) the largest
value $m_t$ of the quark mass such that $Q(m_t)$, and therefore $D(m_t)$, has a
zero eigenvalue. If $m_0^*$ is larger than this value $m_t$ then
$D(m_0^*)=D(m_t)+(m_0^*-m_t) I$ has no negative eigenvalues. Conversely, if
$m_0^*<m_t$, we need to determine the number of zero crossings of the lowest
eigenvalue(s) $\lambda(m_0)$ of $Q(m_0)$ by varying the bare mass $m_0$ from
above $m_t$ down to $m_0^*$. To that end we combine the \texttt{PRIMME}
package with \texttt{openQCD} as mentioned in Ref.~\cite{Mohler:2020txx}.

In practice we proceed in two steps: First we perform a \emph{preselection} to
identify potential candidate configurations with a negative fermion determinant
and for this subset of configurations we perform a \emph{tracking analysis} to
identify the configurations that indeed display a negative fermion determinant.

We start the \emph{preselection} by measuring the lowest O(10) eigenpairs
$(\lambda_i, \psi_i)(m_0^*)$ and their chiralities $\chi_i(m_0^*)$, defined by
\begin{equation}
   \chi_i(m_0^*) = \braket{\psi_i}{\gamma_5 \psi_i}(m_0^*) = \left. \frac{d \lambda_i(m_0)}{d m_0} \right|_{m_0 = m_0^*}\,,
\end{equation}
where the last equality follows from the Feynman-Hellman
theorem~\cite{Akemann:2010em,Mohler:2020txx}. The chirality hence corresponds to
the slope of the eigenvalue function. This allows to categorise the eigenvalues
of $Q$ into those which approach zero as $m_0$ is increased and those which move
away from it. In Figure~\ref{fig:ev-butterfly} we plot the results of the
eigenvalue-chirality analysis for the four lowest lying eigenvalues of the two
$L/a=16$ ensembles with $\kappa=0.1405$ (left) and $\kappa=0.1410$ (right). If a
datapoint falls into the north-east or south-west quadrant, the eigenvalue moves
further away from zero when the quark mass is increased, implying that there is
no zero crossing for values larger than $m_0^*$. This is the case for all configurations with
$\kappa=0.1405$. Conversely, if a datapoint falls into the north-west or
south-east quadrant this implies that the eigenvalue approaches zero as the
quark mass is increased and a zero crossing is possible. Configurations with
eigenvalues which display this feature can potentially have a negative
determinant and therefore require further monitoring. As can be seen in
Fig.~\ref{fig:ev-butterfly}, on the $\kappa=0.1410$ ensembles we find a small
number of these cases for which the second step, the \emph{tracking analysis},
is performed.

We find that datapoints close to the horizontal axis but in the ``safe''
quadrants, tend to occur only for comparably large values of
$\abs{\lambda}$. Under the assumption that the chirality changes slowly in the
range of masses explored, even if the sign of $\chi$ were to change, the
corresponding eigenvalues are not expected to be at risk of changing sign. This
assumption is justified {\it{a posteriori}} in the tracking analysis.

On the configurations that displayed datapoints in the north-west or south-east
quadrants we now measure the lowest 20 eigenpairs for several partially quenched
masses around $m_0^*$. The eigenvalue functions $\lambda_i(m_0)$ and the
eigenbasis $\{\psi_i\}$ are assumed to vary slowly and continuously with
$m_0$. Assuming that the different partially quenched masses are sufficiently
close to each other it is possible to track how a particular eigenvalue behaves
as a function of the quark mass as follows. For each set of neighbouring masses
$m_0$ and $m_0+\Delta m_0$ we construct the matrix $M_{ij}=\braket{\psi_i(m_0)}{
  \psi_j(m_0 + \Delta m_0)}$ of scalar products between the $i$th eigenvector
$\psi_i(m_0)$ at $m_0$ and the $j$th eigenvector $\psi_j(m_0+\Delta m_0)$ at
$m_0+\Delta m_0$. We determine the largest entry $M_{ij}$ and interpret this to
mean that the eigenvalue $i$ at $m_0$ evolves to be the eigenvalue $j$ at
$m_0+\Delta m_0$. We then remove row $i$ and column $j$ from the matrix and
iterate the procedure until each eigenpair at $m_0$ has been assigned a
corresponding eigenpair at $m_0+\Delta m_0$.  Figure \ref{fig:ev-tracking}
displays a configuration of the $L/a=16$ and $\kappa = 0.1410$ ensemble where a
negative determinant was detected. We observe that the line connecting the red
downward facing triangles does cross zero as the mass $m_0$ is increased from
$m_0^*$ (highlighted as the vertical dashed line). Since there is only a
single eigenvalue crossing zero in the region $m_0 > m_0^*$, we conclude that
the fermion determinant is negative on this particular configuration.

We see from the representative example shown in Figure~\ref{fig:ev-tracking}
that the assumption discussed above is indeed valid and the derivatives of the
eigenvalues change very little in the range of masses explored.  We also see
that such derivatives are either of $O(1)$ or small.  This is expected and in
agreement with the discussion in Ref.~\cite{Itoh:1987iy}, where approximate
relations are derived between the eigenpairs corresponding to small eigenvalues
of $Q$ and those of $D$. The chiralities are expected to be significantly
different from zero (and in that case close to $\pm 1$) only for eigenvectors
corresponding to almost real eigenvalues of $D$.

We performed the above analysis for the two smallest values of the quark mass
corresponding to $\kappa = 0.1405$ and $0.1410$ for which we each have a
$L/a=16$ and a $L/a=24$ ensemble. As discussed above (cf. left panel in
Fig.~\ref{fig:ev-butterfly}) we did not observe any cases of a negative
determinant for $\kappa=0.1405$ on either of the two available volumes. Since
negative eigenvalues are expected to have a higher likelihood to occur at small
quark masses, we did not perform this analysis for any of the remaining larger
masses. At $\kappa=0.1410$ we found 6 configurations with a negative determinant
for each of the two volumes.  Furthermore, we observed that the negative sector
is visited at most for the Monte Carlo time corresponding to two consecutive
measurements.  This might be related to our choice of parameters for the
rational approximation of $\sqrt{D^{\dagger}D}$ yielding a relatively low
barrier between the two sectors.  We conclude that in our computational setup
the sign problem for $N_f=1$ QCD is mild and the relative frequency of a
negative determinant of the Dirac matrix is at the sub-percent level.

\section{Correlator analysis \label{sec:cors}}
In order to obtain the spectrum of one-flavour QCD, we create mesonic
correlation functions for states with a variety of quantum numbers. We are
particularly interested in states with scalar (S), pseudoscalar (P) and vector
(I) quantum numbers. We employ the Laplacian Heaviside (LapH)
method~\cite{Morningstar:2011ka,Peardon:2009gh} which allows us to efficiently
compute quark-line disconnected contributions that appear in the computation of
mesonic quantities with a single flavour.

\subsection{Construction of correlation functions}\label{subsec:cors}
Following Ref.~\cite{Morningstar:2011ka} and, where possible, using the same
notation we compute the $N_{v}$ lowest eigenpairs $(\lambda_i, v_i)$ of the
three-dimensional gauge-covariant Laplacian using a stout smeared gauge
field. On each time slice $t$ we arrange these eigenvectors into a matrix $V_s$
as
\begin{equation}
  \begin{aligned}
    V_s(t) = \left(v_1,v_2,\cdots, v_{N_v}\right)
  \end{aligned}
\end{equation}
from which we then define the Hermitian smearing matrix as a function of the
number of eigenpairs that were computed as
\begin{equation}
  \begin{aligned}
    \mc{S}(N_v,t) = V_s (t)V_s^\dagger(t).
  \end{aligned}
\end{equation}
Using a low number of eigenpairs corresponds to a broad smearing profile,
whereas using a large number of eigenpairs corresponds to ``less'' smearing and
taking the limit of all eigenpairs recovers the identity. Quark lines $\mc{Q}$
are computed as
\begin{equation}
  \begin{aligned} \mc{Q}(t_0,t) &= \mc{S}(t) (\gamma_4 D)^{-1} \mc{S}(t_0) \\ &=
    V_s(t) \left[V_s^\dagger(t) (\gamma_4 D)^{-1} V_s(t_0) \right]
    V_s^\dagger(t_0).  \end{aligned}
\end{equation}
The inversion $(\gamma_4 D)^{-1}V_s(t_0)$ is done by solving the equation
\begin{equation}
   (\gamma_4 D)_{\alpha \beta}(t_0,t) y^i_\beta(t) = v_i(t_0)
\end{equation}
for $y^i_\beta(t)$. This is done for each eigenvector $v_i$ ($i=1,\ldots,N_v$),
each spin component ($\alpha=1,\ldots,4$) and each time slice ($t_0 = 0,\ldots,
T-1$), amounting to $N_t\times N_v\times 4$ inversions per configuration.

In our simulation, we keep the number of eigenvalues $N_v=20$ fixed for all
ensembles. However, from these inversions we can construct operators which use
fewer than $20$ eigenvalues by truncating the elements of the square
matrix $V_s^\dagger(t) (\gamma_4 D)^{-1} V_s(t_0)$. Using this we compute meson
correlation functions for $N_v \in \{1,2,3,4,5,6,7,8,9,10,12,15,17,20\}$, which
describe the same spectrum but have different smearing functions.

In all three channels (P, S, I), we use the appropriate interpolation operator
($\mc{P}$, $\mc{S}$, $\mc{I}$) in the finite volume irreducible
representation. For the S--channel we additionally construct a purely gluonic
operator $\mc{G}$~\cite{Brett:2019tzr} which induces the same quantum numbers as
the $\mc{S}$ operator\footnote{To avoid confusion we use the calligraphic
notation for specific operators and Roman letters to indicate the induced
quantum numbers.}. We consider all mutual combinations of $\mc{G}$ and $\mc{S}$
in the 'scalar-glue' system.

\begin{figure}
   \includegraphics[width=\columnwidth]{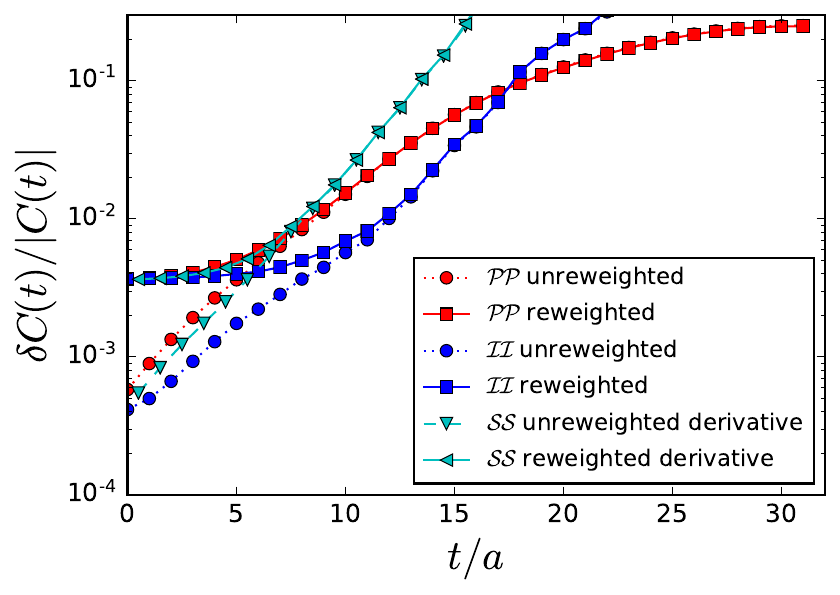}
   \caption{Impact of the reweighting on the relative uncertainties of the
   correlation functions.}
   \label{fig:revs_impact}
\end{figure}

\subsection{Reweighting and vacuum expectation value subtraction}
The vacuum subtracted correlation function $C_{\mc{X}\mc{Y}}$ can be derived
from the un-subtracted correlation function $C^\mathrm{raw}_{\mc{X}\mc{Y}}(t)$
and the vacuum expectation values (vevs) $v_{\mc{X}}$ and $v_{\mc{Y}}$ as
\begin{equation}
   C_{\mc{X}\mc{Y}}(t) = \avg{C^{\mathrm{raw}}_{\mc{X}\mc{Y}}(t)} - \avg{v_{\mc{X}}}\avg{v_{\mc{Y}}}\,,
   \label{eq:vevsub}
\end{equation}
where $\avg{\cdot}$ denotes the gauge average. Whilst the vev is exactly zero
for the $\mc{P}$ operator and numerically zero for the $\mc{I}$ operator, it is
sizable for the $\mc{S}$ and $\mc{G}$ operators. We find that the statistical
signal for correlation functions including $\mc{G}$ or $\mc{S}$ deteriorates
when reweighting (cf. Sec.~\ref{sec:ens}) is combined with the naive vacuum
expectation value subtraction defined in Eq.~\eqref{eq:vevsub}.

This is due to the fact that the product of the vevs is many orders of
magnitude larger than the exponentially decaying part of the correlator and as
a consequence even the little noise introduced by the reweighting factors
destroys the signal for the latter almost completely.

Since the vacuum expectation value is time-independent, an alternative way to
perform the vev subtraction is to take the temporal derivative of the
un-subtracted correlation function. We find that this results in a significantly
better signal when combined with reweighting and are therefore utilising this.

Figure~\ref{fig:revs_impact} displays the effect of reweighting for the example
of the $N_v=20$ correlation functions on the $L/a=20$, $\kappa=0.1390$
ensemble. The figure shows the relative uncertainties of the correlation
function for the $\mc{PP}$ (red), $\mc{II}$ (blue) and the time derivative of
the $\mc{SS}$ (cyan) operators. The dotted lines connect the un-reweighted data
points, whilst the solid lines connect the reweighted ones. We observe that only
for the earliest time slices the uncertainty of the reweighted data is limited
by the accuracy of the reweighting factors.

\begin{figure} 
\includegraphics[width=.48\textwidth]{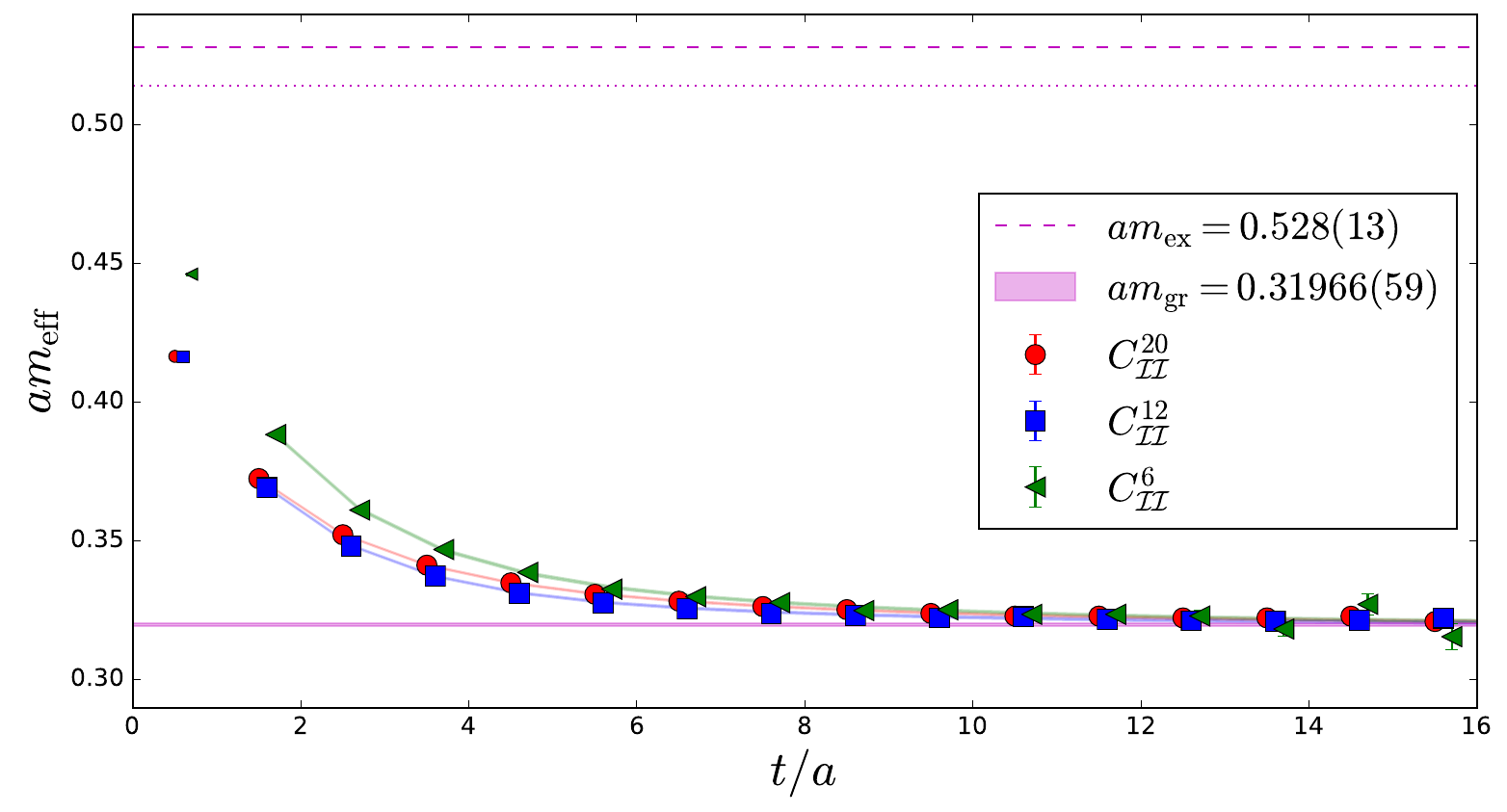}
\includegraphics[width=.48\textwidth]{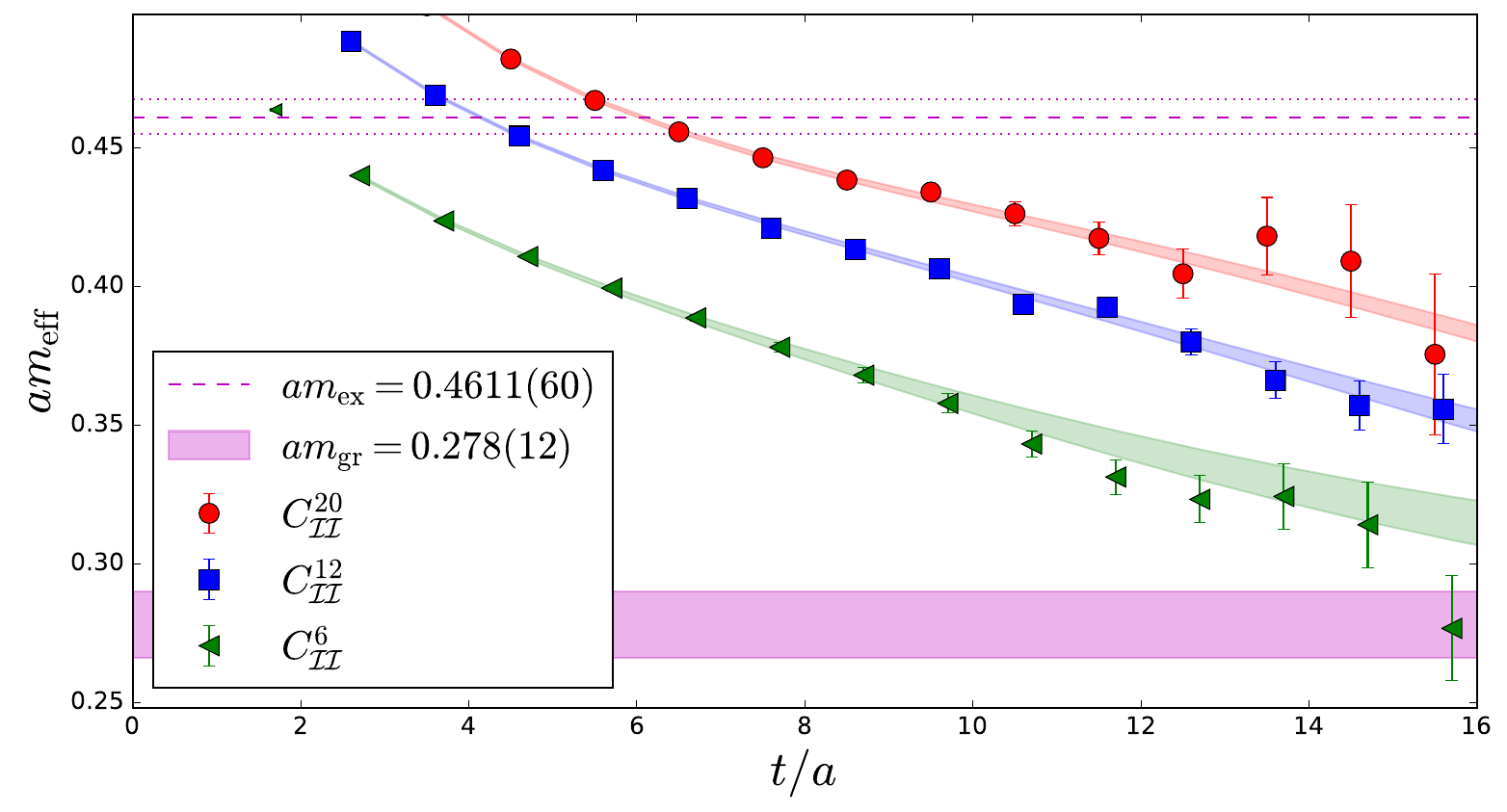}
\caption{Example fit for the vector two point function for the $L/a=32$, $\kappa=0.1400$ (top) and the $L/a=16$, $\kappa=0.1390$ (bottom) ensembles. The datapoints show the effective masses of the underlying correlation functions, whilst the correspondingly coloured bands show the effective mass obtained from the results of the correlation function fits. Finally the magenta horizontal band (dashed line) show the results for the extracted ground (excited) state energies.}
\label{fig:examplecorfit}
\end{figure}

\subsection{Correlation function fits} \label{subsec:fits}
For a given channel (P, I or S), the correlation function $C$ of operators
$O_\mc{X}^n$ with $\mc{X}\in \{\mc{S},\mc{P},\mc{I},\mc{G}\}$ using $n$
eigenvalues can be approximated by the first $N$ states $X_i$ as
\begin{equation}
  \begin{aligned}
    C^n_{\mc{X}\mc{Y}}(t) &= \sum_{i=0}^N \abs{(Z_\mc{X}^n)^*_{i}(Z_\mc{Y}^n)_{i}} \frac{e^{-m^X_i t} + e^{-m^X_i (T-t)}}{2m^X_i}\,,
  \end{aligned}
  \label{eq:2pts}
\end{equation}
where $(Z_\mc{X}^n)_{i} = \matrixel{X_i}{(O_\mc{X}^n)^\dagger}{0}$. We emphasise
that the induced masses $m^X_i$ depend on the channel $X$, rather than the
specific operator $\mc{X}$, in particular all combinations of $\mc{S},\mc{G}$
induce the same spectrum $m_i^S$.

We extract the three lowest--lying states of the spectrum by performing
simultaneous correlated fits to the symmetrised correlation functions
$C_{\mc{X}\mc{Y}}^n(t)$ for several choices of $n$ (between 2 and 4). We
illustrate two such fits for the example of the vector channel in
Fig.~\ref{fig:examplecorfit}. We defer the discussion on the slow approach to
the ground state for the bottom panel to Sec.~\ref{subsec:assignment}.  In order
to assess systematic uncertainties associated with the choice of smearing radii,
we vary which $n$ enter into a particular fit. In particular, for the vector and
pseudoscalar channels we perform three different fits, simultaneously fitting
$N_{v}=(20,12,6), (17,10,3)$ or $(20,15,10,5)$ and labelled `fit1', `fit2' and
`fit3', respectively.\footnote{One of the fit choices of the pseudoscalar meson
on the $L/a=24$, $\kappa=0.1410$ ensemble did not yield an invertible covariance
matrix and was therefore excluded. However, as will be discussed later on, this
ensemble does not enter the final analysis.} For the scalar-glue basis we
simultaneously fit $N_{v}=(20,3)$ or $N_{v}=(17,5)$ (`fit1' and `fit2') but
jointly fitting $C_{\mc{S}\mc{S}}$, $C_{\mc{S}\mc{G}}$ and
$C_{\mc{G}\mc{G}}$. In all cases, we fit three states ($N=2$
in \eqref{eq:2pts}), but only the lowest two potentially enter any subsequent
analysis. We list the numerical results for the lowest two states ('gr' and
'ex', respectively) in Table~\ref{tab:2ptfitres} in
Appendix~\ref{app:corfits}. In all further steps of the analysis we consider all
choices of `fit1', `fit2' and `fit3' to propagate any systematic uncertainties.

Finally, we also compute the connected correlation function for the pseudoscalar
meson, which corresponds to a non-existent state in a $N_f=1$ theory and in the
following is therefore referred to as ``fake pion''. As we will discuss in the
following section, $m_\pi^\mathrm{fake} \to 0$ can be used as a proxy for the
massless limit (see also Ref.~\cite{Francis:2018xjd}).  These correlation functions are
generated from standard point sources and follow the same functional form
as Eq.~\eqref{eq:2pts} with the replacement
$Z_\mc{X}^n \to \matrixel{\pi}{(\bar{q}\gamma_5q)^\dagger}{0}$. For these states
we perform fits with $N=0$ and $N=1$. We note that for both $\kappa=0.1410$
ensembles we expect large finite size effects as $m_\pi^\mathrm{fake} L < 3$ and
therefore discard them from the subsequent analysis.

\section{Analysis of the spectrum \label{sec:results}}
\begin{figure*}
   \includegraphics[width=.49\textwidth]{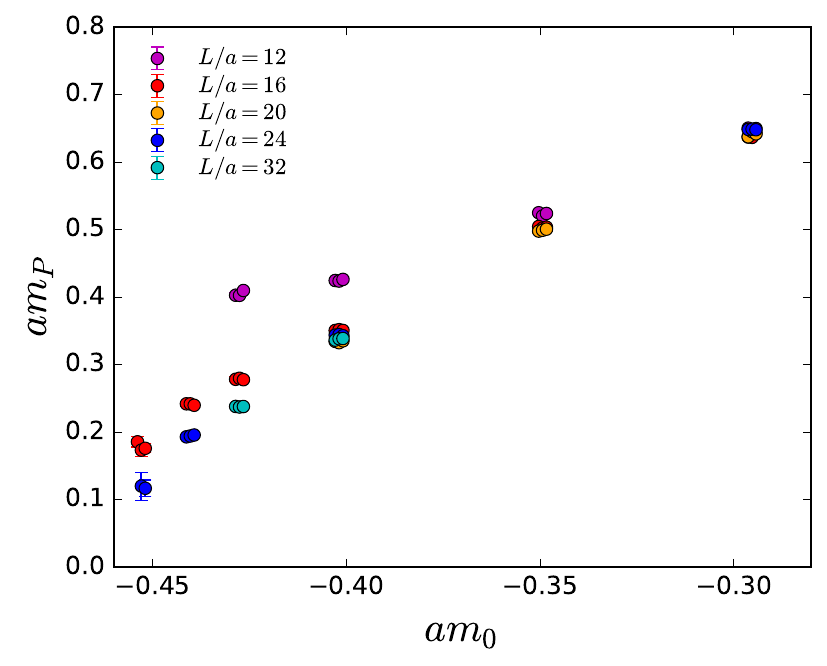}
   \includegraphics[width=.49\textwidth]{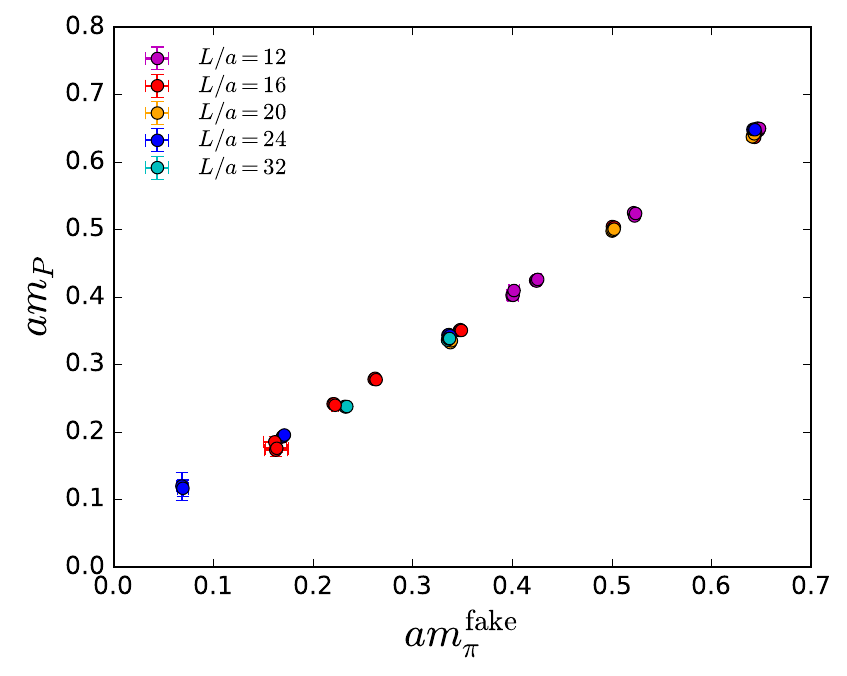}
   \caption{The spectrum of the pseudoscalar meson as a function of the bare
     quark mass (left) and as a function of the fake pion mass (right). Here and in the following,
     shown triplets (or pairs) of points correspond to the fit results of `fit1', `fit2',
     `fit3', respectively.}
\label{fig:P_spectrum}
\end{figure*}

The goal of this section it so extrapolate the results for the meson spectrum
($m_P$, $m_I$ and $m_S$) to the chiral and infinite volume limit to provide
results for ratios of these masses.
\subsection{Defining the chiral limit}\label{subsec:defchiral}

We start by determining what the best proxy for the quark mass
is. Figure~\ref{fig:P_spectrum} shows the lowest lying state for the
pseudoscalar channel. The left panel displays this as a function of the bare
quark mass, the right panel as a function of the fake pion mass. By comparing
the two panels, it is evident that the fake pion mass is the more suitable
choice to define the massless limit as the bare quark mass suffers from large
finite volume effects. Those are due to discretisation effects in the
computation of the critical parameter $\kappa_c$ entering the definition of the
{\it{bare subtracted}}~\footnote{In other words we are saying that data should
  be compared at fixed bare subtracted quark mass and that differs from the bare
  mass by a constant related to $\kappa_c$, which has, at finite lattice
  spacing, a rather strong dependence on the volume~\cite{Sommer:2003ne}.} quark
mass (see Ref.~\cite{Sommer:2003ne} for a discussion in the case of QCD).  In
addition, in Ref.~\cite{Francis:2018xjd} it has been numerically shown, for the
one-flavour $SU(2)$ gauge theory, that the definition of the massless point from
the vanishing of the fake pion mass is consistent with the rigorous definition
from the continuum relation between the topological susceptibility and the quark
mass\footnote{In Ref.~\cite{Leutwyler:1992yt} the relation $\langle \nu^2
  \rangle=\Sigma V m$ is in fact established first for one-flavour QCD and then
  for the case of several flavours. In the equation $\langle \nu^2 \rangle$ is
  the topological susceptibility, $\Sigma$ the fermion condensate and $V$ the
  four-dimensional volume. We see from the plot in Appendix~\ref{app:ensembles}
  that our data are in good qualitative agreement with that relation.}.  In the
following we therefore choose the fake pion mass to define the massless limit.

\begin{figure}
   \includegraphics[width=\columnwidth]{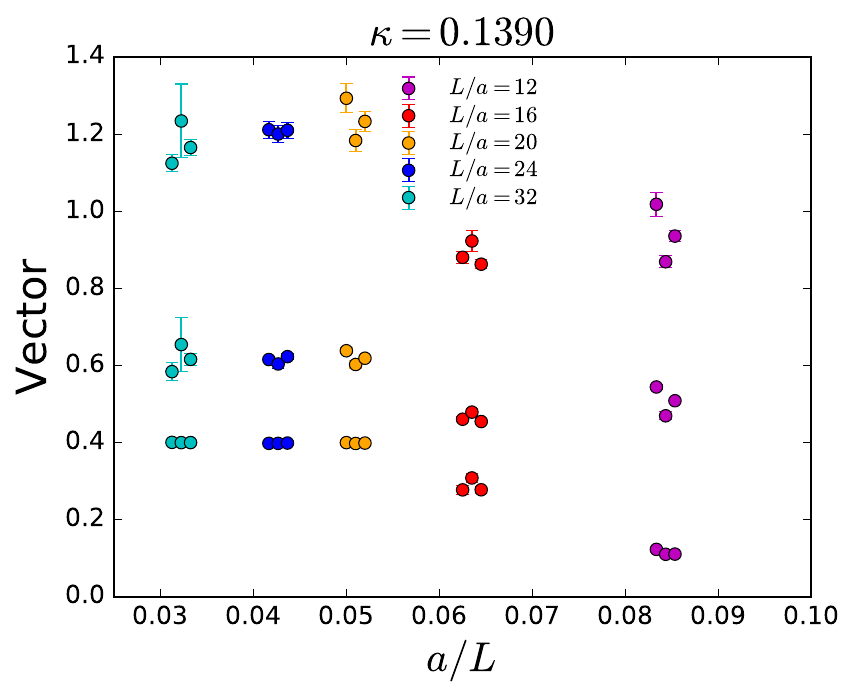}
   \includegraphics[width=\columnwidth]{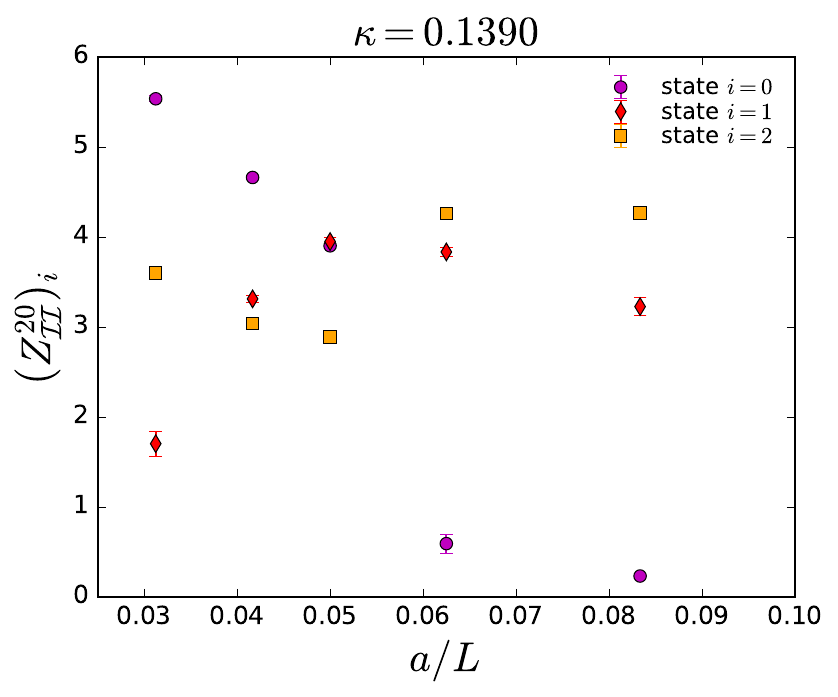}
   \caption{Volume dependence of the vector meson at fixed $\kappa =
     0.1390$. The top panel shows the dependence of the spectrum, the
     bottom panel the dependence of the corresponding matrix elements for
     the $N=20$ correlation function.}
   \label{fig:i_spectrum_voldep}
\end{figure}

\subsection{Assignment of states}\label{subsec:assignment}
To understand the behaviour of the spectrum we induced by means of our chosen
interpolating fields, we investigate how the hadron masses vary as a function of
quark mass and volume. We are predominantly interested in mesonic states
dominated by $q\bar{q}$ contributions\footnote{For the remainder of this work we
  refer to these as ``$q\bar{q}$-states.}. These are expected to display a
strong quark mass dependence but at most a mild dependence on the volume,
whereas any glueball state should only depend weakly on quark mass and
volume. Contrary to these, states that depend mildly on the quark mass but
strongly on the volume do not correspond to physical states and might be
interpreted to be torelon states~\cite{Michael:1986cj,Michael:1990az}.

In Section~\ref{subsec:defchiral} we noted that the pseudoscalar mass is largely
volume independent, but depends smoothly and strongly on the quark mass set by
$m_\pi^\mathrm{fake}$. We therefore identify this with the desired
$q\bar{q}$-state. In the case of the scalar and vector channels, the situation
is more complicated. When comparing results of simulations at the same $\kappa$
but on different volumes, there are cases that display significant volume
dependence on smaller volumes.

\begin{figure}
   \includegraphics[width=\columnwidth]{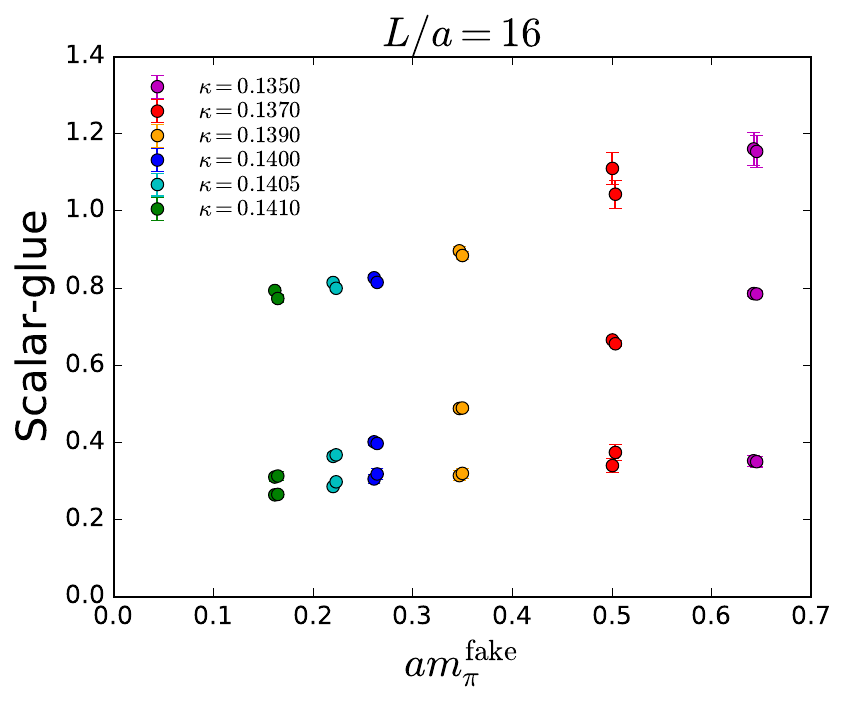}
   \includegraphics[width=\columnwidth]{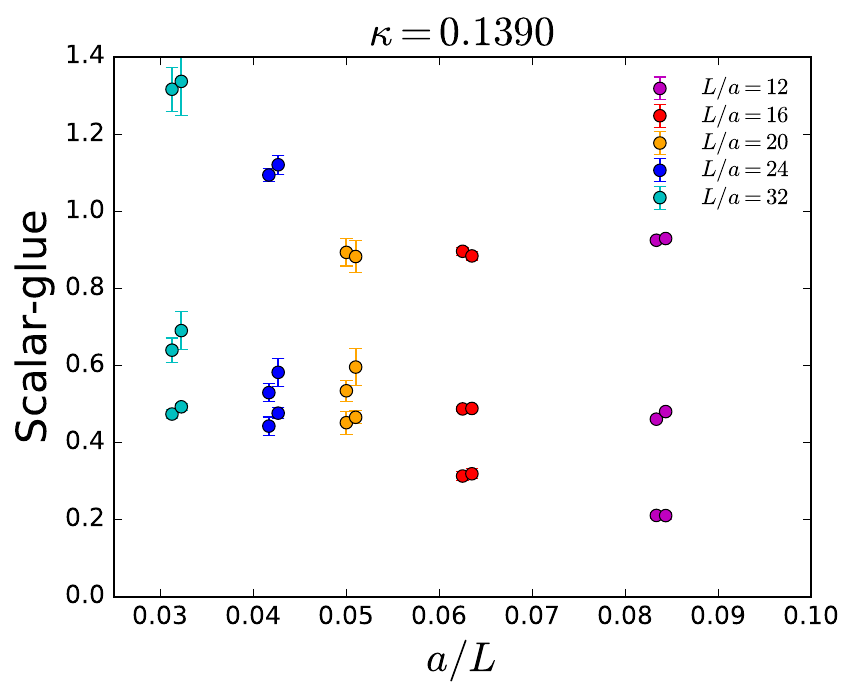}
   \caption{The spectrum of the scalar meson as a function of the quark mass at
     fixed volume $L/a=16$ (top) and as a function of the volume at fixed
     $\kappa = 0.1390$ (bottom).}
   \label{fig:S_spectrum}
\end{figure}

\begin{figure}
   \includegraphics[width=\columnwidth]{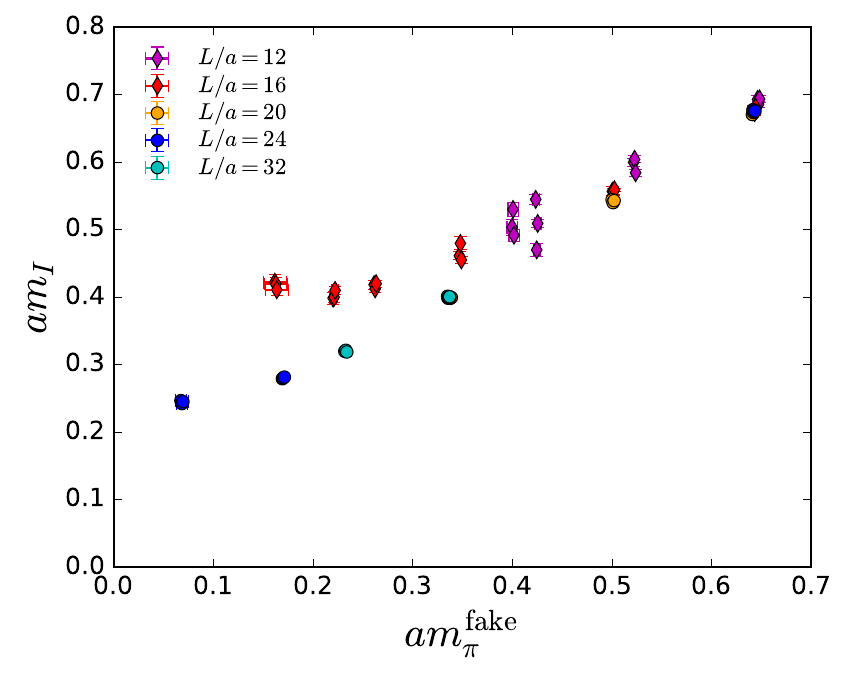}
   \includegraphics[width=\columnwidth]{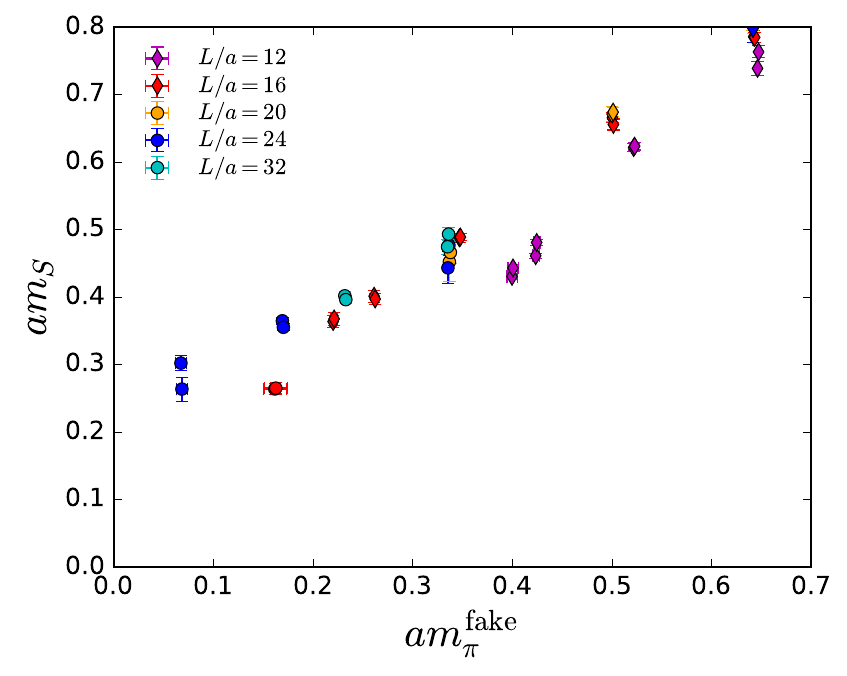}
   \caption{Mass dependence of the states identified as $q\bar{q}$ states for
     the vector (top) and the scalar (bottom).}
   \label{fig:S_i_qq}
\end{figure}

For example, the top panel of Figure \ref{fig:i_spectrum_voldep} shows the
spectrum as a function of the inverse spatial volume but at fixed $\kappa =
0.1390$. We observe that the three largest volumes yield very consistent ground
state masses. Contrary, for the two smallest volumes, we see that a lighter
state is present in the spectrum, which displays a strong volume dependence. We
note that the first excited state on these two volumes is numerically close to
the ground state mass extracted on the larger volumes. This picture is further
substantiated by investigating the behaviour of the amplitude for the matrix
element as we will illustrate with the example of $(Z_\mc{I}^{20})_i$: In the
bottom panel of Figure \ref{fig:i_spectrum_voldep} we show these values for the
three states we are fitting. For the three largest volumes, which are displaying
a consistent ground state mass, we find that the ground state matrix element
(left three magenta circles) is of similar size or larger than the other matrix
elements. In contrast to this, for the smallest two volumes the situation is
reversed and we find the matrix element of the lowest lying state (right two
magenta circles) to be significantly smaller than that of the first and second
excited states. We further note that for these two smallest volumes, the matrix
element corresponding to the first excited state (rightmost two red diamonds)
shows a qualitatively similar behaviour to that of the ground state for the
larger volumes. In other words, for the smallest two volumes, the correlation
function couples more strongly to the first excited state than the ground
state. This is also the reason for the slow approach to the plateau for example
in the case of the $L/a=16$ and $\kappa=0.1390$ ensemble (c.f. bottom panel of
Fig.~\ref{fig:examplecorfit}). The strong volume dependence and qualitatively
different behaviour with respect to the matrix element indicate that the lowest
lying state for the small volumes is not the $q\bar{q}$-state we are interested
in. Instead, as indicated by the values of the mass and the amplitudes we
identify the first excited state with the $q\bar{q}$ state. In summary, for the
vector channel at fixed $\kappa=0.1390$, the $q\bar{q}$ state corresponds to the
lowest lying state for $L/a=32,24,20$ and to the first excited state for
$L/a=16,12$. Corresponding analyses for the other quark masses yield a
similar picture.

Figure \ref{fig:S_spectrum} addresses the scalar channel. The top panel
shows the mass dependence at fixed volume $L/a=16$. The lowest lying state is
mass independent in the range of masses we simulate, but the first excited state
displays a strong mass dependence. The bottom panel shows the volume
dependence at fixed $\kappa=0.1390$. Again, for small volumes, we find a state
whose energy increases as the volume increases (lowest state at $L/a=12,16$), as well as
a volume insensitive state (lowest state at $L/a=32,24,20$ and first excited
state at $L/a=16,12$). Furthermore, the latter coincides with the state that
displayed the strong mass dependence in the top panel. In analogy with the
discussion of the vector meson, we conclude that those correspond to a (mass
dependent, volume independent) scalar meson state and a (mass independent,
volume dependent) torelon state.

By means of similar investigations of the volume and quark mass dependence, we
categorise the two lowest lying states on each ensemble and in each channel into
the lowest quark mass dependent state ($q\bar{q}$) and the remaining state,
which in principle can be a torelon, an excited $q\bar{q}$ or a glueball
state. Figure~\ref{fig:S_i_qq} shows the state that has been identified as the
relevant $q\bar{q}$ state for the vector (top) and scalar (bottom) channels. For
the large volumes, good agreement is found for all quark masses, whereas for
light quark masses and small volumes finite size effects are sizable. We
therefore exclude the $L/a=12$ and $L/a=16$ from our subsequent analysis.

Summarising the discussion in this Section, the $q\bar{q}$ states we are
interested in are easily identified at large volumes and small quark masses as
the lowest lying states in the respective channels. Such determinations have the
largest impact in the chiral and infinite volume extrapolations we discuss next.
However, especially for small volumes, the identification required a more
detailed study of the volume and mass dependence of both the energy levels and
the overlap factors describing the correlation functions.  Those are important
lessons we will take into account for future studies at large values of $N_C$.

\begin{figure}
   \includegraphics[width=\columnwidth]{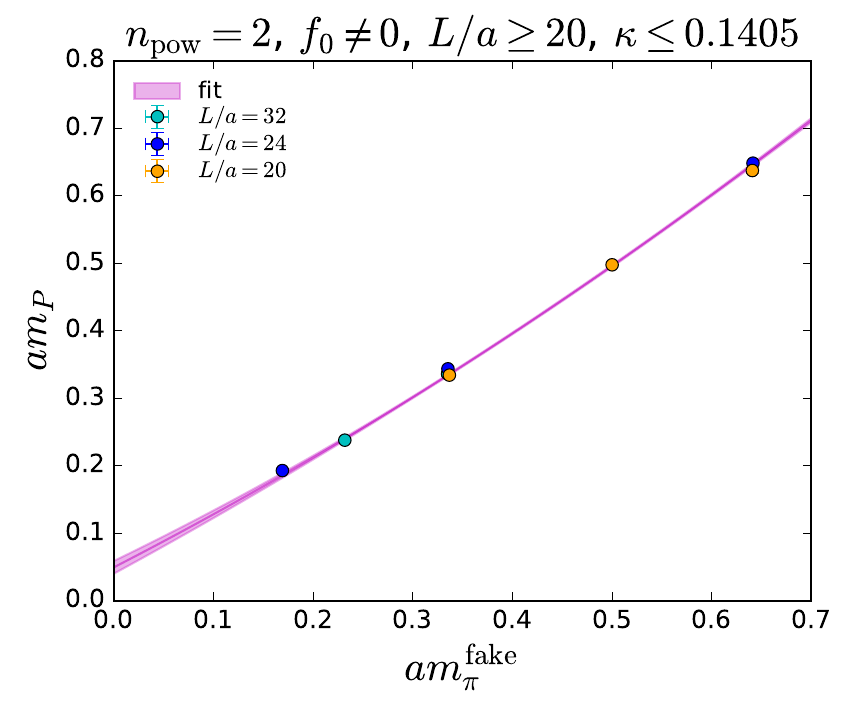}
   \includegraphics[width=\columnwidth]{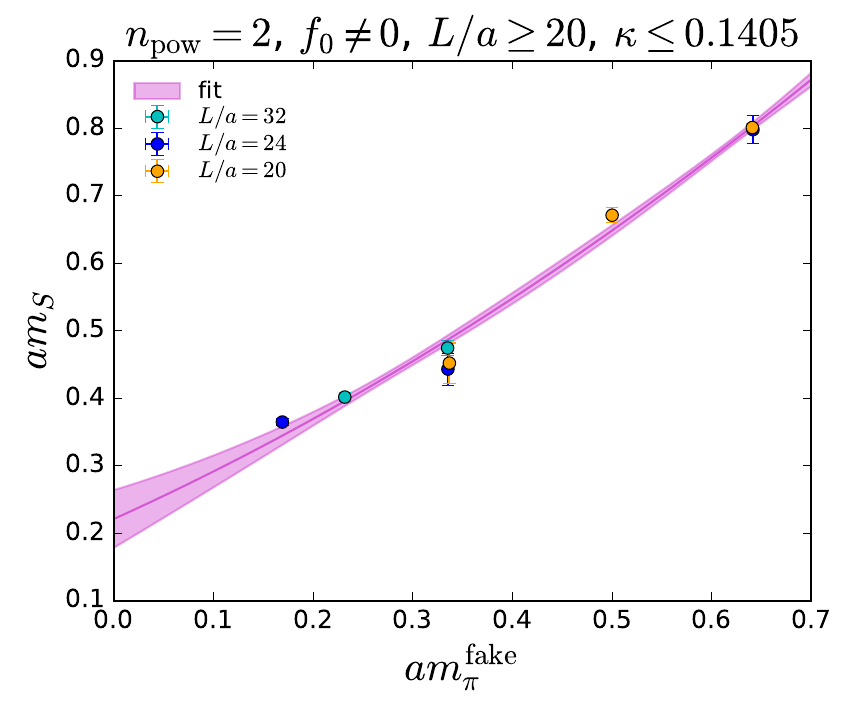}
   \caption{Extrapolation to the chiral limit for a given fit ansatz for the
     pseudoscalar mass (top) and the scalar mass (bottom).}
   \label{fig:chi_fit_P_and_S}
\end{figure}

\begin{figure}
   \includegraphics[width=\columnwidth]{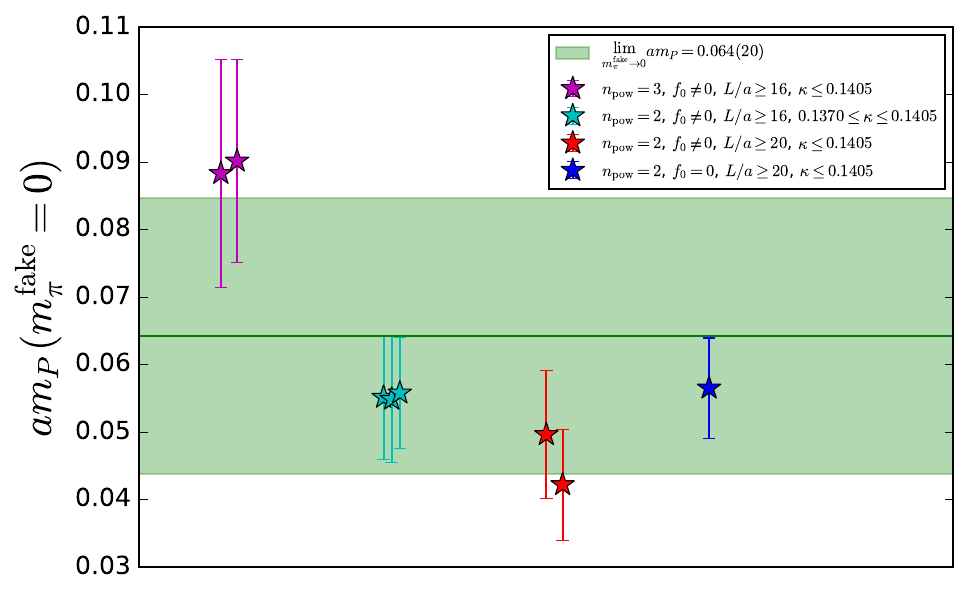}
   \includegraphics[width=\columnwidth]{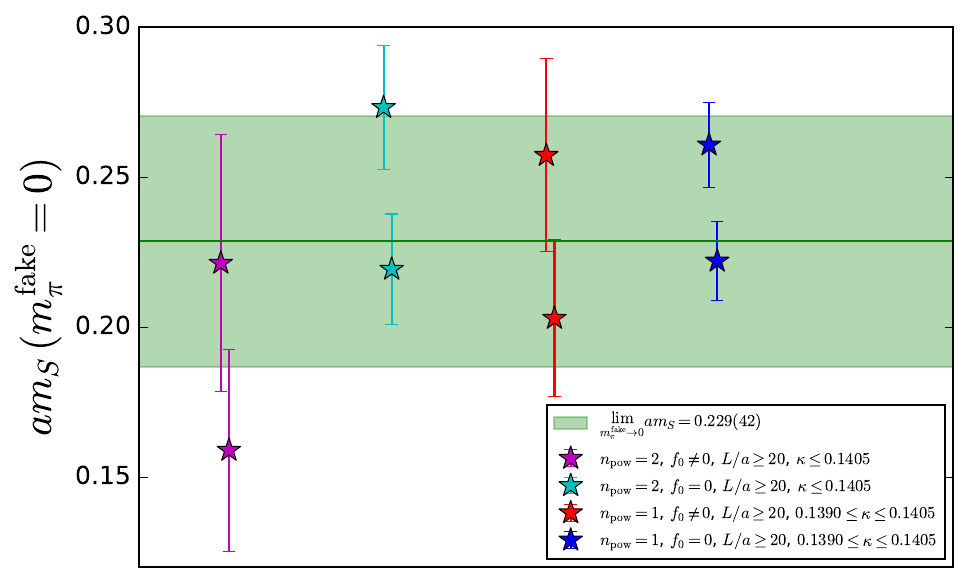}
   \caption{Comparison of the fit results when varying the correlator fit choice
     and the fit ansatz for the pseudoscalar (top) and the scalar mass
     (bottom).}
   \label{fig:fit_comp_P_and_S}
\end{figure}

\begin{figure}
   \includegraphics[width=\columnwidth]{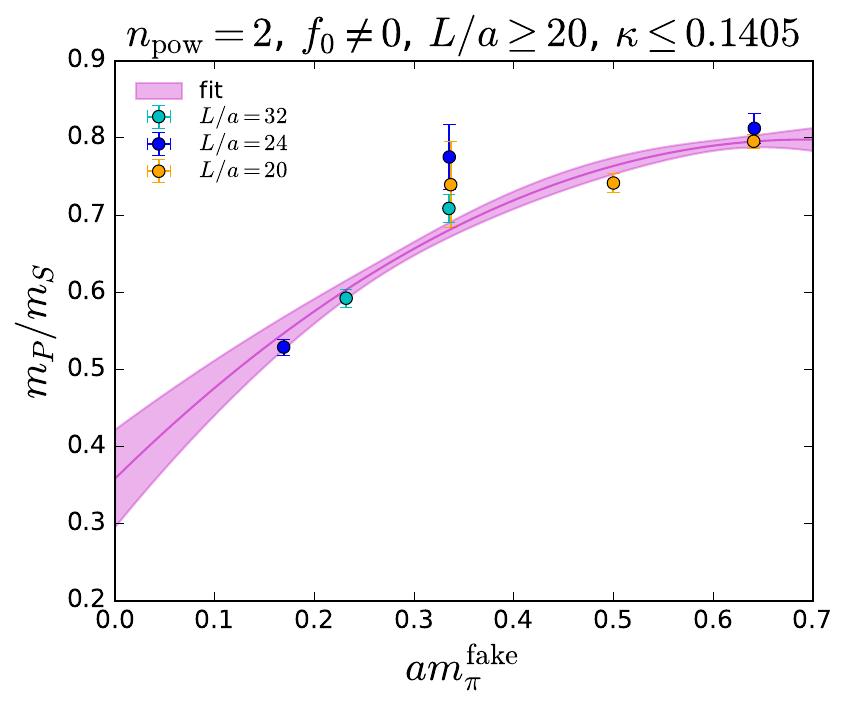}
   \caption{Example extrapolation to the chiral limit of the ratio of pseudoscalar to scalar mass via method~1.}
   \label{fig:chi_fit_RPS}
\end{figure}
\begin{figure*}
  \includegraphics[width=\columnwidth]{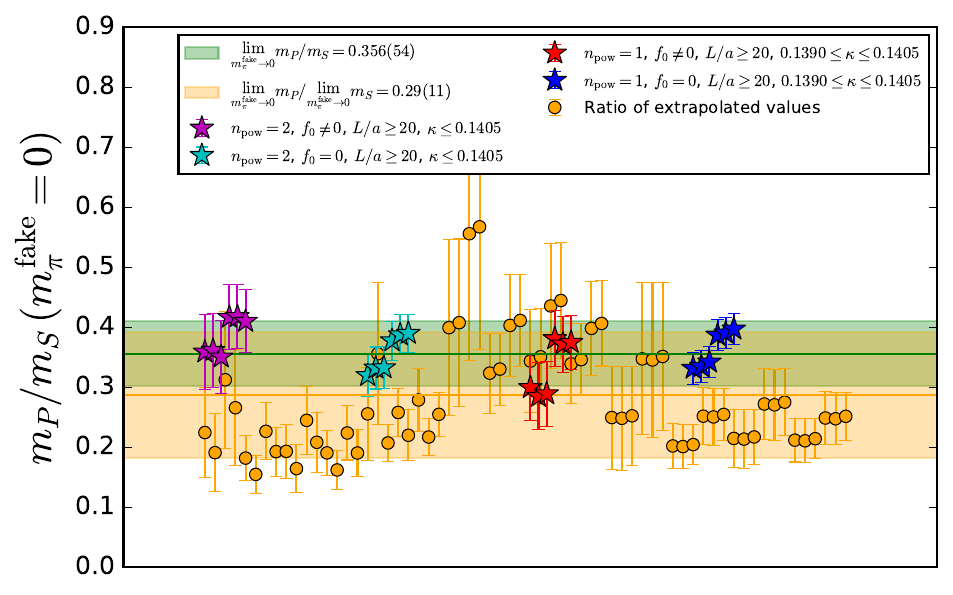}
  \includegraphics[width=\columnwidth]{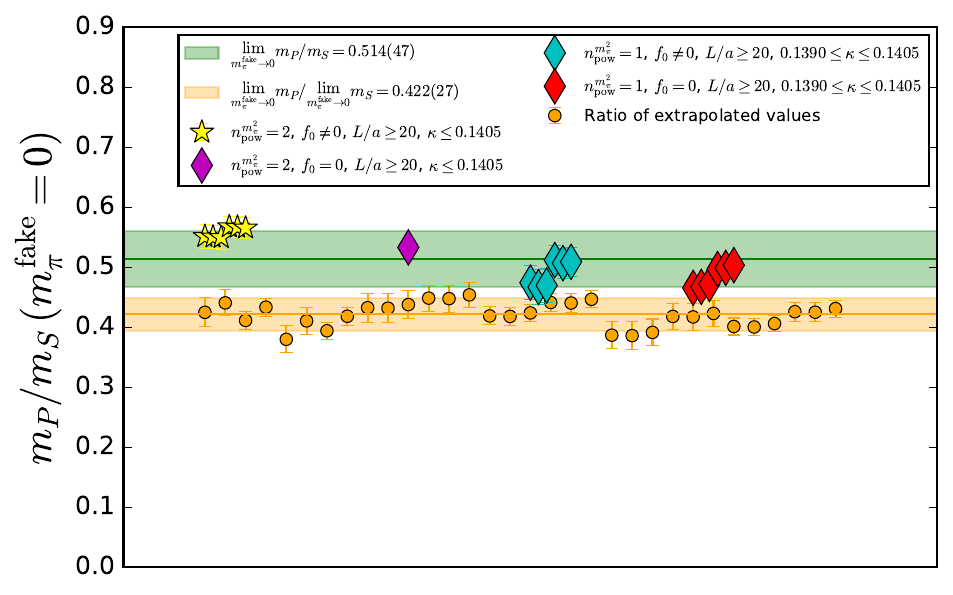}
  \caption{Comparison of fit results for different choices of the extrapolation
    of the ratio of $m_P/m_S$. The left plot shows the extrapolation results
    including even and odd powers in $m_\pi^\mathrm{fake}$ the right plot for
    only even powers.}
   \label{fig:fit_comp_RPS}
\end{figure*}

\subsection{Extrapolation to zero quark mass}
We are interested in the spectrum at vanishing quark mass. Since we have not
performed a scale setting analysis we focus on ratios of masses in the chiral
limit. As discussed above, we will use the fake pion mass to define the zero
quark mass limit.  A completely model independent fit function would have to
include even and odd powers of the fake pion mass. To give a rigorous definition
of the fake pion correlator one would have to consider a partially quenched
theory constructed by introducing a quark field with the same mass parameter as
the original one and quenching it away by a corresponding ghost
field~\cite{DellaMorte:2010aq}. Such a theory would be invariant (at zero quark
mass) under transformations in an extended (graded) chiral symmetry group.
Depending on whether the symmetry is realised \`a la Wigner-Weyl or \`a la
Nambu-Goldstone, one would obtain different relations between the fake pion mass
and the quark mass. In the second case (where the symmetry is broken
spontaneously by the vacuum and explicitly by the quark mass) the quark mass
would turn out to be proportional to the fake pion mass squared.  In this case a
fit in terms of only even powers of the fake pion mass would be more
appropriate.

Any such Gell-Mann-Oakes-Renner-like~\cite{Gell-Mann:1968hlm} relation is valid
at low energies or very close to the massless limit and in the same limit the
fake pion and the pseudoscalar masses should differ significantly, as the first is
expected to vanish while the second not.  Since in our data we only see small
differences between such masses we cannot claim with confidence to be in the
regime where such relations apply. We hence favour the more general fit
function including even and odd powers of the fake pion mass for our final
results. 

\begin{figure}
   \includegraphics[width=\columnwidth]{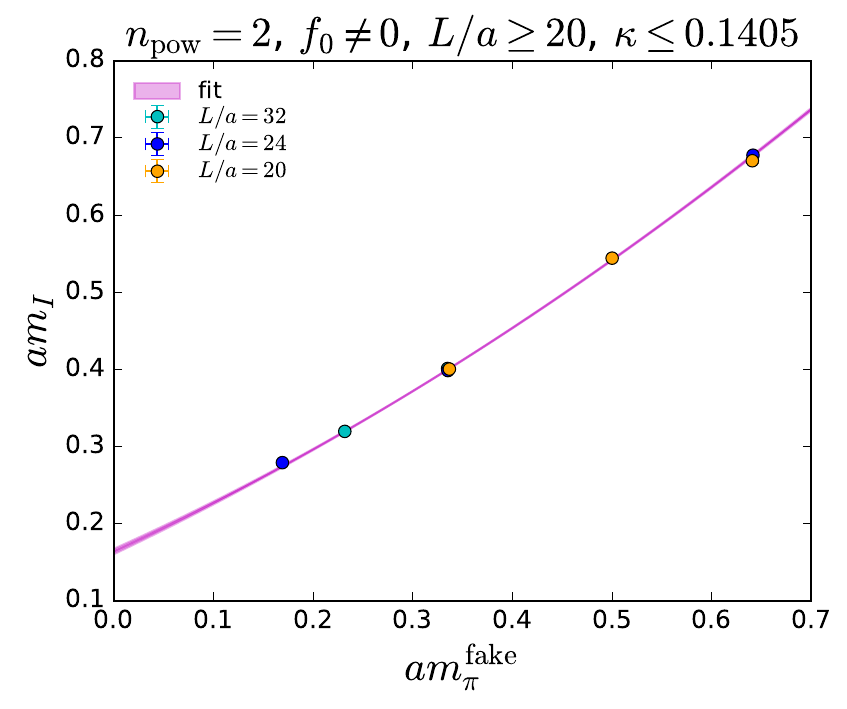}
   \caption{Extrapolation of the vector mass to the chiral
     limit, analogous to Fig.~\ref{fig:chi_fit_P_and_S}
     for different choices.}
   \label{fig:chi_fit_i}
\end{figure}

\begin{figure}
  \includegraphics[width=\columnwidth]{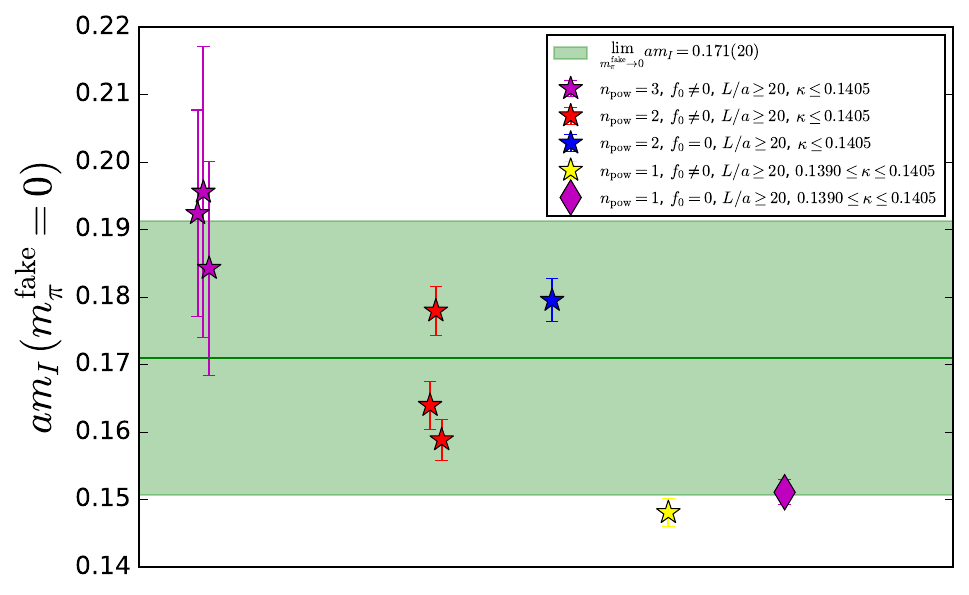}
   \caption{Variations of the extrapolation of the vector mass to the chiral
     limit, analogous to Fig.~\ref{fig:fit_comp_P_and_S}.}
   \label{fig:fit_comp_i}
\end{figure}

\begin{figure}
   \includegraphics[width=\columnwidth]{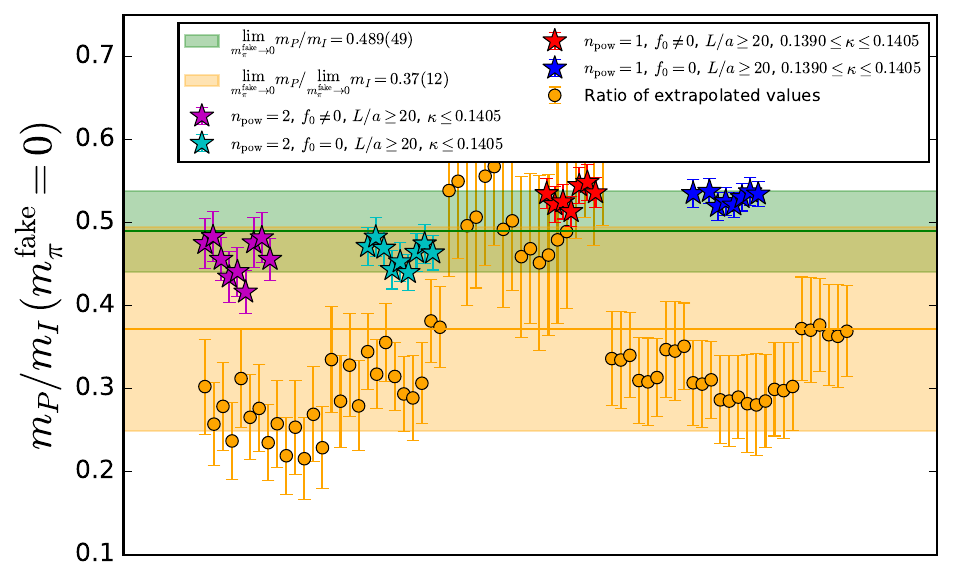}
  \includegraphics[width=\columnwidth]{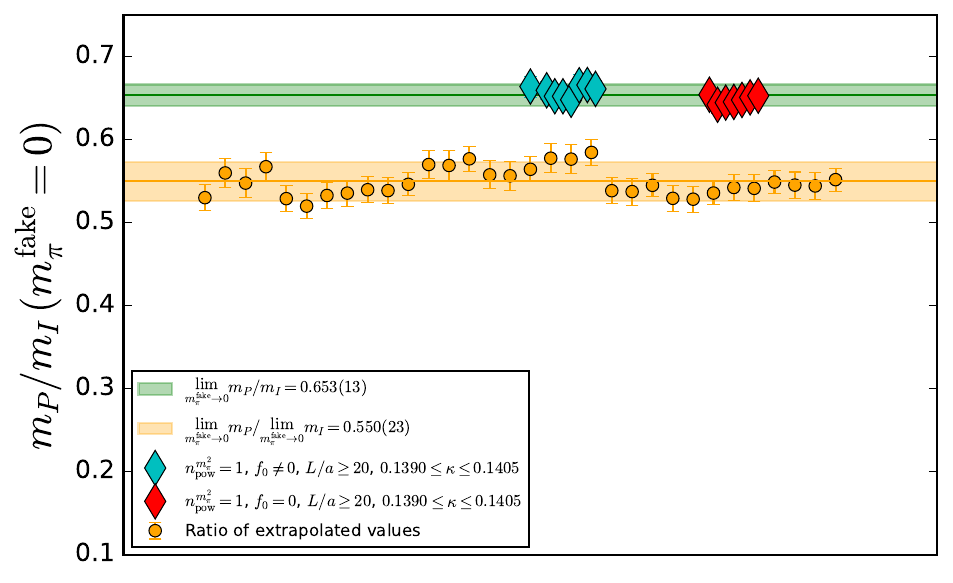}
   \caption{Variations of the extrapolation of the ratio of the pseudoscalar and
     the vector mass using even and odd powers of $m_\pi^\mathrm{fake}$ (top)
     and only even powers of $m_\pi^\mathrm{fake}$ (bottom).}
   \label{fig:fit_comp_RPi}
\end{figure}

\begin{figure}
  \includegraphics[width=\columnwidth]{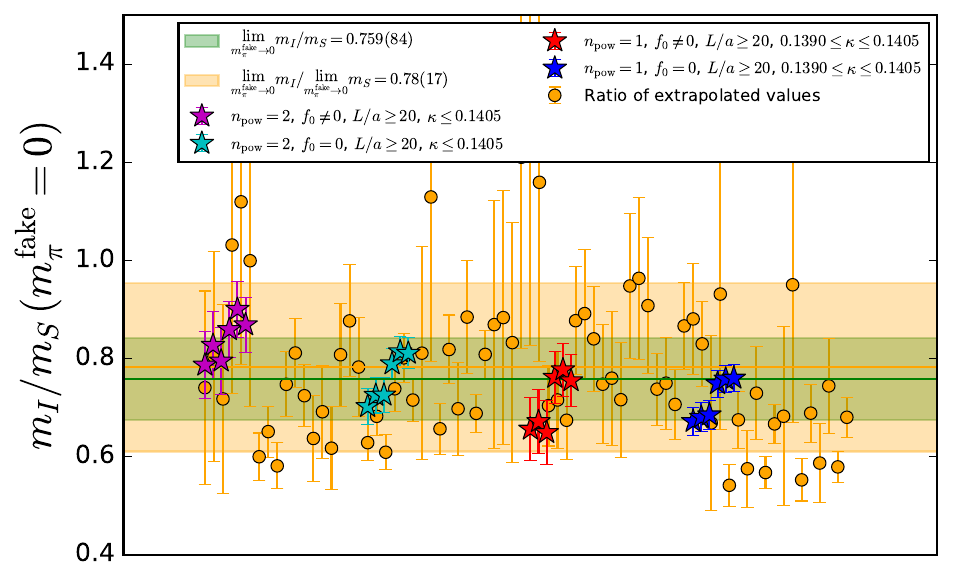}
  \caption{Results of fits to the ratio of the vector and the scalar mass.}
  \label{fig:fit_comp_RiS}
\end{figure}

The fit functions we explore for this extrapolation are
\begin{equation}
   M(m_\pi^\mathrm{fake},L) = \left[\sum_{i=0}^\mathrm{n_\mathrm{pow}} c_i  \left(m^\mathrm{fake}_\pi\right)^i\right] \left(1+ f_0 e^{-m_P L}\right)\,,
\end{equation}
where $M$ is either a mass ($m_P$, $m_S$, $m_I$) or ratios thereof.  In line
with what we discussed above, we separately consider choices where $i$ takes
even and odd values or only even values and in both cases either leaving $f_0$
as a free parameter or setting it to zero.  In addition to varying the fit
function, we consider cuts to the data, in particular removing the smallest
volumes and/or the lightest and/or heaviest masses.

An example fit for the case of the pseudoscalar mass (top) and the scalar mass
(bottom) is shown in Figure~\ref{fig:chi_fit_P_and_S}. In both of these cases we
take the results obtained by `fit1', keep $f_0$ as a free parameter and choose
$n_\mathrm{pow}=2$. Due to concerns about the finite volume effects, we exclude
the smallest volumes ($L/a=12, 16$) and the lightest quark mass
($\kappa=0.1410$).

We repeat all extrapolations for the various choices of the correlation
function fits, whether or not $f_0$ is kept as a free parameter and for
different choices of $n_\mathrm{pow}$. For the lowest order polynomial we
restrict the mass range that enters the fit. The datapoints in
Fig.~\ref{fig:fit_comp_P_and_S} show the results for these variations for the
pseudoscalar (top) and the scalar (bottom). Only fits with an acceptable
$p$-value of $p>0.05$ are shown. The green band in these plots is derived by
taking the 68th percentile of the distribution of the underlying bootstrap
samples of all the fits which produced an acceptable $p$-value. We interpret
this number to be a good approximation of systematic effects due to correlator
fit choices, variations of the chiral fit ansatz and the data included in
such a fit.

Ultimately we are interested in the ratio of masses in the chiral limit. We can
obtain this in two ways as we will now illustrate on the example of the ratio of
the pseudoscalar to the scalar mass: We can either build the ratio $m_P/m_S$ at
finite $m_\pi^\mathrm{fake}$ and then extrapolate this to the massless limit
(method~1), or we can separately extrapolate the pseudoscalar and the scalar
masses and then build their ratio (method~2). One example fit of the former is
shown in Fig.~\ref{fig:chi_fit_RPS}. We observe that part of the mass dependence
cancels in the ratios, resulting in a less steep curve than that observed in the
individual fits (cf. Fig~\ref{fig:chi_fit_P_and_S}). The coloured stars
in the left panel of Fig.~\ref{fig:fit_comp_RPS} show different
variations of the fit ansatz, analogous to Fig.~\ref{fig:fit_comp_P_and_S}. In
addition to the extrapolation of the ratio of masses (method~1), we also show
ratios of the chirally extrapolated values (orange circles; method~2). Here we
computed all mutual combinations of acceptable fits displayed in
Fig.~\ref{fig:fit_comp_P_and_S}. The green (orange) band is the result of taking
the 68th percentile of all the bootstrap samples for the fits of method~1
(method~2) that produced an acceptable $p$-value.

In general, we notice that the ratio of separate chiral extrapolations leads to
larger variations than the extrapolation of the ratio of masses. This is
unsurprising as, ensemble by ensemble, the underlying datapoints are
statistically correlated, and therefore statistical fluctuations are reduced for
the individual ratios of datapoints. Furthermore the extrapolation of the
individual datapoints is more difficult to control since the slope with the fake
pion mass is steeper. Our preferred number is therefore the direct extrapolation
(green band in Fig.~\ref{fig:fit_comp_RPS}) whilst the orange band provides a
sanity check.

We are now in a position to compare the results of the fits including even and
odd powers of $m_\pi^\mathrm{fake}$ to those only using even powers. These two
choices are compared in the two panels of Fig.~\ref{fig:fit_comp_RPS} for the
example of the ratio of pseudoscalar to scalar masses. The green bands of the
two panels are in $\sim 2 \sigma$ agreement, lending confidence in the
results. However the errorbands of the direct and indirect methods do not
overlap for the fit of the even powers only. This is even more pronounced for
the case of the ratio $m_P/m_I$ (c.f.~the bottom panel of
Fig~\ref{fig:fit_comp_RPi}). This numerical evidence further supports our
preference for the more conservative fit ansatz including even and odd powers of
$m_\pi^\mathrm{fake}$ and we therefore quote results from this choice as our
final numbers.

In addition to $m_P$, $m_S$ we have data for the vector mass $m_I$. An example
fit for the extrapolation of the vector mass is shown in
Fig.~\ref{fig:chi_fit_i} (cf. Fig.~\ref{fig:chi_fit_P_and_S}) whilst different
fit variations are shown in Fig.~\ref{fig:fit_comp_i}
(cf. Fig.~\ref{fig:fit_comp_P_and_S}). Finally, we also construct the ratios
$m_P/m_I$ (see Fig.~\ref{fig:fit_comp_RPi}) and $m_I/m_S$ (see
Fig.~\ref{fig:fit_comp_RiS}) in the chiral limit via the two methods described
above.

\section{Discussion and Outlook \label{sec:discussion}}
We have presented a detailed study of the spectrum of one-flavour QCD using
Wilson fermions with tree-level O$(a)$ improvement.

Results are obtained at one single lattice spacing (approximatively $0.06$ fm)
for different volumes (up to $32^3 \times 64$) and several quark masses.  After
extrapolating to the massless limit we
obtain
\begin{equation}
\frac{m_{P}}{m_S}= 0.356(54) \;,
\end{equation}
for the pseudoscalar to scalar meson mass ratio
and
\begin{equation}
\frac{m_{P}}{m_{I}}= 0.489(49) \;,
\end{equation}
for the pseudoscalar to vector ratio. In Ref.~\cite{Sannino:2003xe} a
prediction using an effective field theory approach and a $1/N_C$ expansion was derived.
In the massless limit this reads
\begin{equation}
\frac{m_{P}}{m_S}= 1-\frac{22}{9N_C} - \frac{4}{9}\beta
+O\left( \frac{1}{N_C^2} \right)\;,
\end{equation}
where $\beta$ is a positive constant of order $1/N_C$.  The equation above
therefore provides an estimate for an upper bound, that for $N_C=3$ reads
\begin{equation}
\frac{m_{P}}{m_S} \lesssim 0.185\; ,
\end{equation}
up to higher order effects starting at $1/N_C^2$.

Our results are somewhat larger than this bound, but considering their
uncertainty and terms of size $O(1/N_C^2)$, they are reasonably close. This
might indicate that $1/N_C^2$ corrections and the parameter $\beta$ are small.
Obviously this finding needs to be corroborated by extending our studies to
larger values of $N_C$.

Those are further motivated, firstly, by the observation about the slope in the
mass for the extrapolation of $m_P/m_S$ having the opposite sign compared to the
prediction in Ref.~\cite{Sannino:2003xe}. Corrections to that start at
$O(1/N_C)$ and can therefore be quite large.  Secondly, results at larger values
of $N_C$ will allow assessing the range of validity of the two different
theoretical predictions in
Refs.~\cite{Sannino:2003xe}~and~\cite{Armoni:2005qr}. The latter predicts a
value of $1/3$ for the ratio $m_P/m_S$ at $N_C=3$ in the massless limit.

We have provided an improved estimate, concerning cutoff effects and assessment
of systematic errors, compared to previous results that appeared as Proceedings
in Ref.~\cite{Farchioni:2008na} (based on Ref.~\cite{Farchioni:2007dw}), where a
value of 0.410(41) was found for the pseudoscalar to scalar mass ratio.

Besides having tested the predictions made in Refs.~\cite{Sannino:2003xe,Armoni:2005qr} for the
spin-zero one flavour QCD mesonic state, we further provided information on the
vector spectrum that can be interpreted as the leading order prediction for the
$\mathcal{N} = 1$ super Yang-Mills vector states.

In order to assess the size of higher order effects we are extending the
computation considering $N_C=4,5$ and 6. A preliminary account appeared in
Ref.~\cite{DellaMorte:2022htz}.

\section{Acknowledgements \label{sec:acknowledgements}}
We thank John Bulava for discussions and his early work.  We thank the members
of the SDU lattice group for useful discussions. The project leading to this
application has received funding from the European Union's Horizon 2020 research
and innovation programme under the Marie Sk{\l}odowska-Curie grant agreement No
894103 and by the Independent Research Fund Denmark, Research Project 1, grant
number 8021-00122B.  F.P.G.Z.~acknowledges support from UKRI Future Leader
Fellowship MR/T019956/1.  This work was partially supported by DeiC National HPC
(g.a. DeiC-SDU-N5-202200006). Part of the computation done for this project was
performed on the UCloud interactive HPC system and ABACUS2.0 supercomputer,
which is managed by the eScience Center at the University of Southern Denmark.

\appendix
\section{Distribution of the topological charge \label{app:ensembles}}
\begin{figure*}
  \includegraphics[width=.95\textwidth]{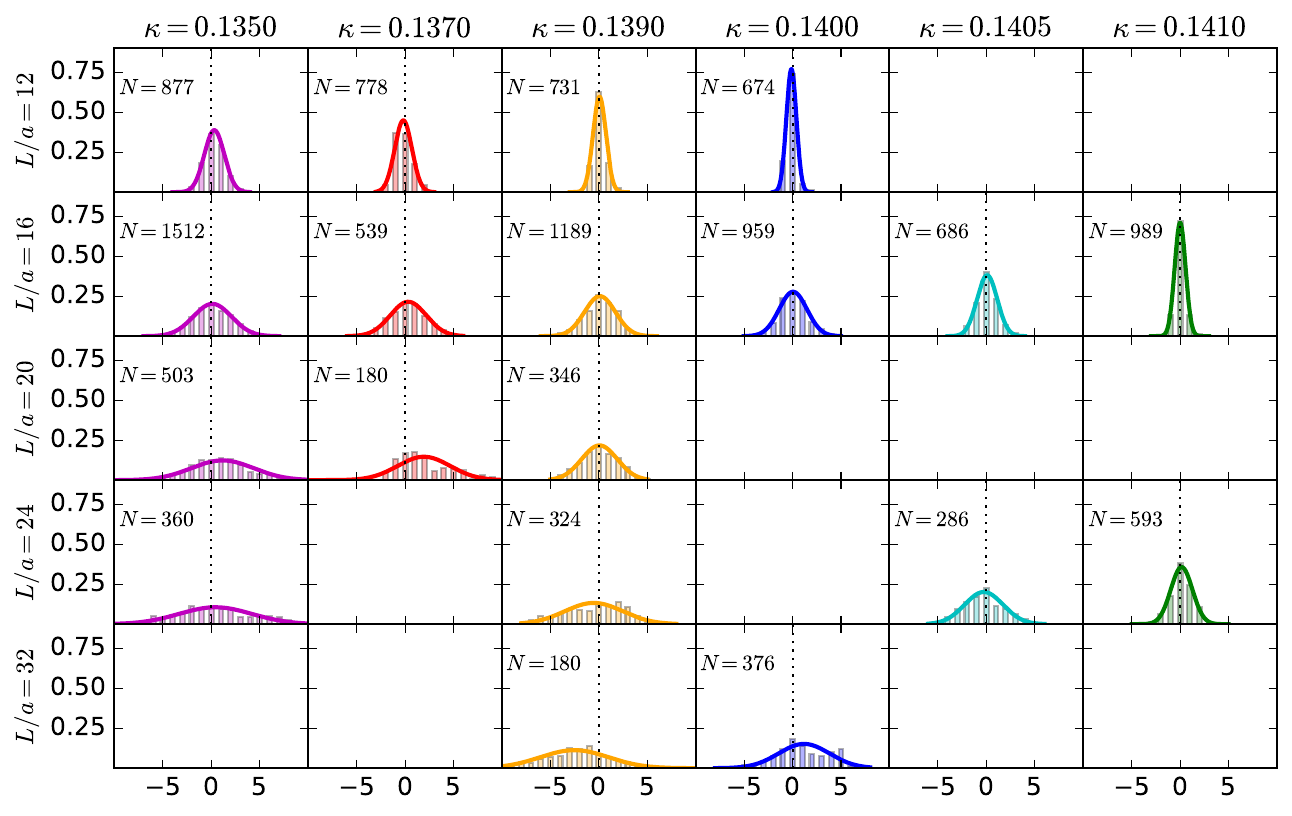}
  \caption{Topological charge distribution for the various ensembles.}
  \label{fig:topo}
\end{figure*}

In Figure~\ref{fig:topo} we show the normalised distributions of the topological
charge on all ensembles. The number $N$ corresponds to the number of distinct
configuration on which all measurements have been performed and which are spaced
by a minimum of 32 trajectories (cf.~Sec~\ref{sec:ens}). We clearly observe that the
  topological charge becomes more peaked as the volume is decreased and as the
  quark mass is lowered (larger values of $\kappa$)~\cite{Leutwyler:1992yt}.
\section{Results of the correlation function fits \label{app:corfits}}
Table~\ref{tab:2ptfitres} shows the relevant results obtained by fitting the
reweighted and vacuum-subtracted correlation functions as described in
Sec~\ref{sec:cors}.

\begin{table*}
\begin{tabular}{lll|lll|lll|ll|l}
\hline
$L/a$ & $\kappa$ & & \multicolumn{3}{c|}{Pseudoscalar} & \multicolumn{3}{c|}{Vector} & \multicolumn{2}{c|}{Scalar} & $m_\pi^\mathrm{fake}$\\ \hline
&  &  & fit1 & fit2 & fit3 & fit1 & fit2 & fit3 & fit1 & fit2 &  \\
\multirow{4}{*}{12}& \multirow{2}{*}{0.1350} & gr &  0.6503(24) &  0.6468(18) &  0.6497(23) &  0.1944(90) &  0.1800(68) &  0.1884(82) &  0.2248(70) &  0.2189(68) & \multirow{2}{*}{ 0.6457(13)} \\
&  & ex &  0.859(20) &  0.787(29) &  0.859(16) &  0.6929(54) &  0.6856(43) &  0.6936(58) &  0.739(10) &  0.7632(88) &  \\
& \multirow{2}{*}{0.1370} & gr &  0.5250(23) &  0.5202(28) &  0.5239(23) &  0.1500(66) &  0.1663(71) &  0.1367(53) &  0.1987(76) &  0.1957(81) & \multirow{2}{*}{ 0.5207(21)} \\
&  & ex &  0.794(12) &  0.737(15) &  0.7796(99) &  0.5998(60) &  0.6041(52) &  0.5842(55) &  0.6215(62) &  0.6236(57) &  \\
& \multirow{2}{*}{0.1390} & gr &  0.4246(28) &  0.4238(30) &  0.4262(27) &  0.1237(41) &  0.1110(42) &  0.1115(33) &  0.2117(84) &  0.2113(89) & \multirow{2}{*}{ 0.4229(36)} \\
&  & ex &  0.6845(87) &  0.667(11) &  0.7019(71) &  0.5447(76) &  0.470(10) &  0.5091(61) &  0.4613(37) &  0.4809(39) &  \\
& \multirow{2}{*}{0.1400} & gr &  0.4028(35) &  0.4026(39) &  0.4097(34) &  0.1081(39) &  0.1209(48) &  0.1101(31) &  0.272(21) &  0.283(23) & \multirow{2}{*}{ 0.4008(45)} \\
&  & ex &  0.6912(91) &  0.683(14) &  0.7198(87) &  0.503(12) &  0.530(11) &  0.4916(86) &  0.4307(50) &  0.4432(58) &  \\
\hline
\hline
\multirow{6}{*}{16}& \multirow{2}{*}{0.1350} & gr &  0.6371(25) &  0.6367(31) &  0.64301(92) &  0.540(47) &  0.370(43) &  0.507(39) &  0.352(14) &  0.350(14) & \multirow{2}{*}{ 0.64189(36)} \\
&  & ex &  0.901(31) &  0.859(34) &  0.832(16) &  0.6765(70) &  0.6734(39) &  0.6809(19) &  0.7858(88) &  0.7848(78) &  \\
& \multirow{2}{*}{0.1370} & gr &  0.5046(21) &  0.5023(23) &  0.5036(17) &  0.356(41) &  0.358(29) &  0.396(31) &  0.340(17) &  0.374(22) & \multirow{2}{*}{ 0.50102(84)} \\
&  & ex &  0.763(26) &  0.786(35) &  0.794(18) &  0.5568(66) &  0.5561(48) &  0.5590(24) &  0.6655(74) &  0.6557(81) &  \\
& \multirow{2}{*}{0.1390} & gr &  0.3506(21) &  0.3517(21) &  0.3506(20) &  0.278(12) &  0.309(11) &  0.2782(96) &  0.314(12) &  0.320(13) & \multirow{2}{*}{ 0.34695(93)} \\
&  & ex &  0.619(20) &  0.618(31) &  0.621(18) &  0.4611(60) &  0.4796(96) &  0.4551(46) &  0.4879(65) &  0.4890(55) &  \\
& \multirow{2}{*}{0.1400} & gr &  0.2782(34) &  0.2798(35) &  0.2777(33) &  0.2118(76) &  0.2134(69) &  0.2203(60) &  0.305(11) &  0.318(13) & \multirow{2}{*}{ 0.2623(17)} \\
&  & ex &  0.510(18) &  0.514(25) &  0.516(16) &  0.4186(66) &  0.4122(56) &  0.4199(53) &  0.4014(88) &  0.3972(84) &  \\
& \multirow{2}{*}{0.1405} & gr &  0.2419(36) &  0.2418(46) &  0.2398(36) &  0.1768(74) &  0.1819(70) &  0.1872(58) &  0.2856(56) &  0.2977(77) & \multirow{2}{*}{ 0.2239(38)} \\
&  & ex &  0.554(18) &  0.563(32) &  0.554(15) &  0.3985(90) &  0.4000(79) &  0.4100(62) &  0.3638(93) &  0.368(10) &  \\
& \multirow{2}{*}{0.1410} & gr &  0.1856(76) &  0.173(10) &  0.1760(65) &  0.1595(78) &  0.1520(61) &  0.1554(66) &  0.2639(78) &  0.2651(87) & \multirow{2}{*}{ 0.162(12)} \\
&  & ex &  0.468(18) &  0.427(25) &  0.462(14) &  0.422(11) &  0.4194(99) &  0.4108(89) &  0.310(10) &  0.313(11) &  \\
\hline
\hline
\multirow{3}{*}{20}& \multirow{2}{*}{0.1350} & gr &  0.6375(50) &  0.6451(46) &  0.6419(39) &  0.6705(38) &  0.6735(18) &  0.6746(21) &  0.484(37) &  0.500(40) & \multirow{2}{*}{ 0.64106(34)} \\
&  & ex &  0.815(36) &  0.802(43) &  0.805(76) &  0.86(11) &  0.936(40) &  0.854(32) &  0.8011(58) &  0.7992(89) &  \\
& \multirow{2}{*}{0.1370} & gr &  0.4978(42) &  0.4990(42) &  0.5006(35) &  0.5442(17) &  0.5400(25) &  0.5430(18) &  0.309(51) &  0.356(68) & \multirow{2}{*}{ 0.49905(77)} \\
&  & ex &  0.783(32) &  0.777(21) &  0.766(24) &  0.737(19) &  0.722(24) &  0.711(19) &  0.671(11) &  0.6741(64) &  \\
& \multirow{2}{*}{0.1390} & gr &  0.3344(13) &  0.3325(15) &  0.3353(14) &  0.4004(10) &  0.3982(14) &  0.3993(10) &  0.452(30) &  0.466(17) & \multirow{2}{*}{ 0.33608(80)} \\
&  & ex &  0.6757(92) &  0.6686(70) &  0.6683(90) &  0.6387(73) &  0.6030(92) &  0.6193(70) &  0.535(28) &  0.596(49) &  \\
\hline
\hline
\multirow{4}{*}{24}& \multirow{2}{*}{0.1350} & gr &  0.6484(25) &  0.6488(15) &  0.6480(22) &  0.6777(15) &  0.6759(11) &  0.6757(12) &  0.564(92) &  0.71(13) & \multirow{2}{*}{ 0.64141(26)} \\
&  & ex &  0.862(25) &  0.852(36) &  0.866(24) &  0.861(29) &  0.933(25) &  0.873(24) &  0.798(20) &  0.834(28) &  \\
& \multirow{2}{*}{0.1390} & gr &  0.3437(26) &  0.3445(34) &  0.3428(25) &  0.39849(67) &  0.39851(81) &  0.39934(58) &  0.443(24) &  0.477(15) & \multirow{2}{*}{ 0.33522(45)} \\
&  & ex &  0.601(18) &  0.596(24) &  0.602(18) &  0.6159(68) &  0.604(11) &  0.6234(68) &  0.530(23) &  0.583(36) &  \\
& \multirow{2}{*}{0.1405} & gr &  0.1930(30) &  0.1942(33) &  0.1957(29) &  0.2791(15) &  0.2804(16) &  0.2815(14) &  0.3649(51) &  0.3553(51) & \multirow{2}{*}{ 0.1691(14)} \\
&  & ex &  0.525(10) &  0.513(12) &  0.515(11) &  0.5108(78) &  0.5078(86) &  0.5172(73) &  0.623(42) &  0.571(40) &  \\
& \multirow{2}{*}{0.1410} & gr & - &  0.120(21) &  0.117(12) &  0.2466(28) &  0.2423(28) &  0.2451(27) &  0.302(11) &  0.264(18) & \multirow{2}{*}{ 0.0673(57)} \\
&  & ex & - &  0.501(70) &  0.468(58) &  0.503(11) &  0.491(13) &  0.498(11) &  0.598(42) &  0.559(52) &  \\
\hline
\hline
\multirow{2}{*}{32}& \multirow{2}{*}{0.1390} & gr &  0.3363(38) &  0.3383(34) &  0.3387(31) &  0.40116(57) &  0.4004(17) &  0.40054(42) &  0.475(11) &  0.4933(93) & \multirow{2}{*}{ 0.33484(42)} \\
&  & ex &  0.562(24) &  0.556(34) &  0.592(21) &  0.585(23) &  0.655(69) &  0.616(16) &  0.640(32) &  0.691(49) &  \\
& \multirow{2}{*}{0.1400} & gr &  0.2380(22) &  0.2372(22) &  0.2379(17) &  0.31966(59) &  0.32121(45) &  0.31876(50) &  0.4019(71) &  0.3963(63) & \multirow{2}{*}{ 0.23229(39)} \\
&  & ex &  0.541(18) &  0.549(19) &  0.553(12) &  0.528(13) &  0.640(20) &  0.5240(98) &  0.571(34) &  0.564(26) &  \\
\hline
\end{tabular}

\caption{Fit results from correlation function fits for the pseudoscalar, vector
  and scalar channels as well as the fake pion mass. Only statistical
  uncertainties are quoted. \label{tab:2ptfitres}}
\end{table*}

\section{Determination of $t_0$ \label{app:t0}}

\begin{figure}
  \includegraphics[width=\columnwidth]{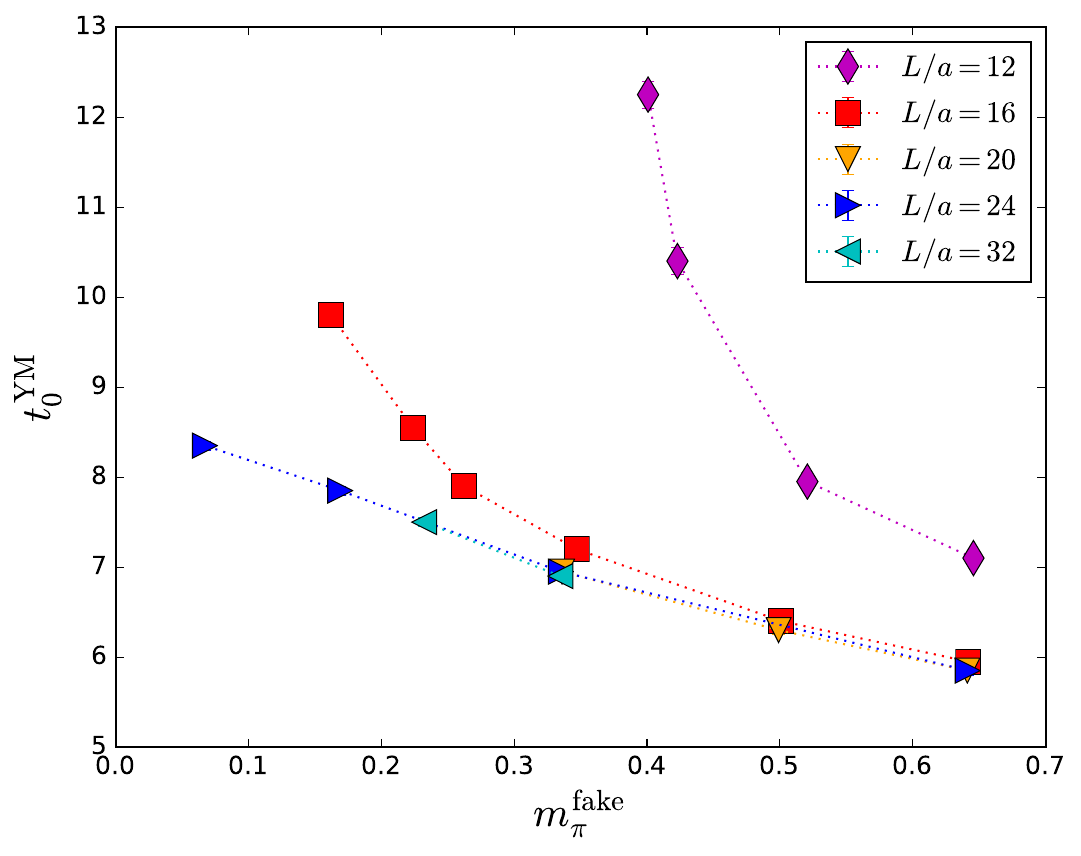}
  \caption{Results for the $t_0^{\mathrm{YM}}$ using the Wilson
    flow~\cite{Luscher:2010iy} as a function of fake pion mass. The equivalent
    plot for $t_0^{\mathrm{Wilson}}$ displays the same qualitative features.
    \label{fig:t0}}
\end{figure}

Table~\ref{tab:t0} shows values obtained for $t_0$ on each ensemble. We use two
different action densities to compare systematic effects of setting the scale,
i.e. the Wilson plaquette action ($t^{\mathrm{Wilson}}_0$) and the Yang-Mills action
($t^{\mathrm{YM}}_0 $). As mentioned before, we quote these values only as reference for
other lattice simulations. The quoted uncertainties originate from the numerical
integration scheme and are statistical only. As an example we show the
dependence of $t^{\mathrm{YM}}_0$ on the fake pion mass in Fig.~\ref{fig:t0}. We note
that $t_0^{\mathrm{YM}}$ displays large finite size effects for the smallest ensembles.
\begin{table}
  \begin{tabular}{cccccc}
    \hline
    $L/a$ & $\kappa$ & $t^{\mathrm{Wilson}}_0/a^2$ & $t^{\mathrm{YM}}_0/a^2$ & $a^{\mathrm{Wilson}}$ & $a^{\mathrm{YM}}$\\
    \hline\hline
    12 & 0.1350 & 7.40(8) & 7.10(5) & 0.058(1)& 0.060(1)\\
    12 & 0.1370 & 8.25(8) & 7.95(5) & 0.055(1)& 0.056(1)\\
    12 & 0.1390 & 10.75(8) & 10.40(15) & 0.049(1)& 0.049(1)\\
    12 & 0.1400 & 12.55(8) & 12.25(15) & 0.045(1)& 0.045(1)\\
    \hline
    16 & 0.1350 & 6.15(5) & 5.95(5) & 0.064(1)& 0.065(1)\\
    16 & 0.1370 & 6.60(5) & 6.40(5) & 0.062(1)& 0.063(1)\\
    16 & 0.1390 & 7.40(5) & 7.20(5) & 0.058(1)& 0.059(1)\\
    16 & 0.1400 & 8.10(5) & 7.90(5) & 0.056(1)& 0.057(1)\\
    16 & 0.1405 & 8.80(5) & 8.55(5) & 0.054(1)& 0.054(1)\\
    16 & 0.1410 & 10.10(5) & 9.80(5) & 0.050(1)& 0.051(1)\\
    \hline
    20 & 0.1350 & 6.10(5) & 5.85(5) & 0.064(1)& 0.066(1)\\
    20 & 0.1370 & 6.50(5) & 6.30(5) & 0.062(1)& 0.063(1)\\
    20 & 0.1390 & 7.15(5) & 6.95(5) & 0.059(1)& 0.060(1)\\
    \hline
    24 & 0.1350 & 6.05(5) & 5.85(5) & 0.065(1)& 0.066(1)\\
    24 & 0.1390 & 7.15(5) & 6.95(5) & 0.059(1)& 0.060(1)\\
    24 & 0.1405 & 8.05(5) & 7.85(5) & 0.056(1)& 0.057(1)\\
    24 & 0.1410 & 8.60(5) & 8.35(5) & 0.054(1)& 0.055(1)\\
    \hline 
    32 & 0.1390 & 7.10(5) & 6.90(5) & 0.060(1)& 0.061(1)\\
    32 & 0.1400 & 7.75(5) & 7.50(5) & 0.057(1)& 0.058(1)\\
    \hline
    \end{tabular}

  \caption{Results for the $t_0$ using the Wilson flow~\cite{Luscher:2010iy}. 
   The quoted uncertainties are statistical only and discrete by construction, as 
   the Wilson flow is numerically integrated and saved every step of length 0.05. We quote 
   two estimates using the Wilson plaquette action ($t^{\mathrm{Wilson}}_0$) and 
   the Yang-Mills action ($t^{\mathrm{YM}}_0 $). The difference can be seen as 
   an estimate for systematic uncertainties of the overall procedure. \label{tab:t0}}
\end{table}

\bibliography{oneflavour.bib}

\end{document}